%% file: main.tex
\RequirePackage{fix-cm}
\documentclass[smallextended,nospthms]{svjour3}

\smartqed  

\usepackage{packages}  
\usepackage{orcidlink}

\begin{document}
\journalname{Empirical Software Engineering}

\title{\revised{Green Architectural Tactics in ML-enabled Systems: An LLM-based Repository Mining Study}}

\titlerunning{\revised{Green Tactics in ML-enabled Systems}}      

\author{Vincenzo De Martino \orcidlink{0000-0003-1485-4560} \and
        Silverio Martínez-Fernández \orcidlink{0000-0001-9928-133X} \and
        Fabio Palomba \orcidlink{0000-0001-9337-5116}}

\institute{
Vincenzo De Martino \and Fabio Palomba \at
Software Engineering (SeSa) Lab, University of Salerno, Fisciano, Italy \\
\email{\{vdemartino, fpalomba\}@unisa.it}
\and
Vincenzo De Martino \and Silverio Martínez-Fernández \at
Universitat Politècnica de Catalunya (UPC), Barcelona, Spain \\
\email{\{vincenzo.de.martino, silverio.martinez\}@upc.edu}
}


\date{Received: date / Accepted: date}  

\maketitle

\begin{abstract}
\textbf{Context:} The increasing adoption of machine learning (ML) and artificial intelligence (AI) technologies raises growing concerns about their environmental sustainability. Developing and deploying ML-enabled systems is computationally intensive, particularly during training and inference. Green AI has emerged to address these issues by promoting efficiency without sacrificing accuracy. While prior research has proposed catalogs of sustainable practices (i.e., green tactics), there remains limited understanding of their adoption in practice and whether additional, undocumented tactics exist. \textbf{Objective:} This study aims to investigate the extent to which existing sustainable practices are implemented in real-world ML-enabled systems and to identify previously undocumented practices that support environmental sustainability. \textbf{Method:} We conduct a mining software repository study on \revised{205} open-source ML projects on \textsc{GitHub}. To support our analysis, we design a novel mechanism based on large language models (LLMs) capable of identifying both known and new sustainable practices from code repositories. \textbf{Results:} Our findings confirm that green tactics reported in the literature are used in practice, although adoption rates vary. Furthermore, our LLM-based approach reveals nine previously undocumented sustainable practices. Each tactic is supported with code examples to aid adoption and integration. \textbf{Conclusions:} We finally provide insights for practitioners seeking to reduce the environmental impact of ML-enabled systems and offer a foundation for future research in automating the detection and adoption of sustainable practices.

\end{abstract}

\input{1_Introduction}
\input{2_RelatedWorks}
\input{3_Methodology}

\input{5_Result}

\input{6_Discussion}
\input{7_Conclusion}

\section*{Acknowledgments}
ChatGPT was used to improve the language and readability of this work. The authors reviewed and edited the content and take full responsibility for it. This work has been partially supported by the Grant PID2024-156019OB-I00 funded by MICIU/AEI/10.13039/501100011033 by ERDF, EU, and \textsl{GAISSA-Optimizer} research project funded by the AGAUR agency (Code: 2025 PROD 00236). This work has been partially supported by the European Union - NextGenerationEU through the Italian Ministry of University and Research, Projects PRIN 2022 "QualAI: Continuous Quality Improvement of AI-based Systems" (grant n. 2022B3BP5S, CUP: H53D23003510006).

\section*{Data availability statement} The manuscript includes data as supplementary material. In particular, the original dataset of ML projects, preprocessed data, and all scripts used for data collection, preparation, and analysis are useful for reproducing the study. The material is available as part of our online appendix on Figshare \cite{appendix}.

\section*{Declaration}

\textbf{Conflict of Interest} The authors declare that they have no known competing financial interests or personal relationships that could have appeared to influence the work reported in this paper.

\section*{Author Contributions}
\textbf{Vincenzo De Martino}: Conceptualization, Formal analysis, Investigation, Data Curation, Methodology, Validation, Visualization, Writing - Original Draft, Writing - Review \& Editing.
\textbf{Silverio Martínez-Fernández}: Conceptualization, Supervision, Visualization, Writing - Review \& Editing.
\textbf{Fabio Palomba}: Conceptualization, Visualization, Supervision, Resources, Writing - Review \& Editing.

\bibliographystyle{spmpsci}  
\bibliography{bib}

\end{document}

%% file: 1_Introduction.tex
\section{Introduction}
\label{sec:introduction}
The widespread influence of machine learning (ML) and artificial intelligence (AI) in various sectors - such as education, medicine, science, and finance - has been well documented, and these technologies drive significant gains in innovation and efficiency \cite{wu2022sustainable}. As a subset of AI, ML focuses on creating algorithms that can learn from data and make predictions or decisions without being explicitly programmed \cite{sarker2021machine}. Systems incorporating AI or ML, often referred to as AI-enabled or ML-enabled systems, are becoming increasingly common \cite{martinez2022software}. The growing use of these systems raises concerns about their sustainability, particularly in terms of energy consumption and CO\textsubscript{2} emissions, which have become increasingly pressing \cite{lacoste2019quantifying,duran2024identifying,de2025sustainability}. The development and deployment of ML-enabled systems are resource-intensive, especially during training and inference \cite{schwartz2020green}. \revised{For example, training models like LLaMA-65B consume around 449 MWh of electricity, producing approximately 173 tons of CO\textsubscript{2} emissions \cite{touvron2023llama}, which corresponds to powering an average electric car for approximately 5,058 km (3,143 miles)\cite{zhao2023quantifying}, raising concerns about their long-term sustainability, especially considering the increasing demand for powerful models.}

While the focus has traditionally been on maximizing accuracy, there is increasing recognition of the need to balance this with environmental impact, leading to the concept of Green AI \cite{schwartz2020green}. Green AI advocates for reducing computational demands while maintaining accuracy, emphasizing a responsible approach to AI development and deployment \cite{van2021sustainable}.
Maintaining this balance is essential to promote sustainable ML engineering and is important for society and users of ML systems, who are increasingly concerned about climate change. By addressing the environmental impact of ML-enabled systems, Green AI helps foster trust and align technological progress with broader societal goals of environmental responsibility and climate action \cite{martinez2024environmental}. 

In recent years, the software engineering community has made considerable efforts in promoting software sustainability, with specific attention now being paid to ML-enabled systems \cite{verdecchia2023systematic,tamburri2020sustainable,martinez2023towards}. This shift has led to the identification and synthesis of various green tactics aimed at reducing the environmental footprint of ML models \cite{verdecchia2023systematic,10.1145/3639475.3640111}. For instance, Järvenpää et al. \cite{10.1145/3639475.3640111} recently elicited a comprehensive catalog of 30 green architectural tactics for ML-enabled systems, which were derived from a thorough analysis of the literature and validated through expert focus groups. These tactics encompass strategies across different phases of ML development and deployment, such as data-centric and management. A code example of a green architectural tactic is the quantization techniques provided by PyTorch\footnote{\url{https://pytorch.org/docs/stable/quantization.html}}, such as dynamic quantization shown below, to optimize ML models.
\begin{lstlisting}
import torch 
# create a model instance
model_fp32 = M()
# create a quantized model instance
model_int8 = torch.ao.quantization.quantize_dynamic(
    model_fp32,  # the original model
    {torch.nn.Linear},  # a set of layers to dynamically quantize
    dtype=torch.qint8)  # the target dtype for quantized weights
\end{lstlisting}

While the work by Järvenpää et al. \cite{10.1145/3639475.3640111} offers valuable guidance on Green AI, the actual relevance of these tactics is not yet established. 

\steattentionboxa{\faWarning \hspace{0.05cm}
There is still a gap of knowledge on the extent to which green tactics are actually implemented in real-world contexts. Moreover, it is unclear if there are additional practices not yet documented in the literature that practitioners employ when developing ML-enabled systems.}

An improved understanding of the practical adoption of green AI tactics is crucial for both researchers and practitioners. Researchers can gain insights into the relevance and impact of certain tactics, which can inform future studies and refinements of existing guidelines. Practitioners can benefit from knowledge of the most widely employed tactics in the wild, which can guide them in making sustainable choices. Furthermore, assessing the impact of the documented tactics will make previous findings more actionable and tangible, thereby enhancing their practical utility.


To overcome the limitations of previous studies, we recently presented a preliminary short paper \cite{de2024developers} where we investigated the extent to which developers adopt green tactics defined in the literature when developing ML-enabled systems. Specifically, we build upon the catalog proposed by Järvenpää et al. \cite{10.1145/3639475.3640111}, analyzing the adoption of these tactics in 168 open-source ML projects on \textsc{GitHub}. To this aim, we developed a novel LLM-based mechanism to identify green architectural tactics in software repositories at the code level. This earlier work revealed that while developers implement many cataloged tactics, others remain underutilized due to knowledge gaps or implementation challenges. In this article, we build on this foundation by addressing two critical limitations of the earlier study. 

\begin{itemize}
    \item First, \textbf{we broaden the scope beyond the predefined catalog of tactics} by enhancing the LLM-based mechanism to identify undocumented green tactics employed by developers in real-world ML-enabled systems. This approach allowed us to \textbf{extend the existing catalog by introducing new tactics that were not previously documented}.

    \smallskip
    \item \minor{Second, we augment the empirical findings of the preliminary study with additional \textbf{quantitative and qualitative insights supported by detailed examples of the newly identified tactics}, and expanding our dataset from \textbf{168} to \textbf{205 open-source ML projects}}.

    \smallskip
    \item Furthermore, \textbf{we provide an extended and refined description of the methodology employed to investigate the adoption of green tactics}, enhancing the comprehensiveness of our analysis. 

    \smallskip
    \item Finally, we provided discussions and implications of our work for researchers and practitioners, \textbf{highlighting how the findings of our study may impact future developments in green software engineering}.
\end{itemize} 

Our results reveal that all existing tactics in the literature are adopted in real software projects, with a different frequency for each tactic. Additionally, starting by analyzing code rather than papers as previous work \cite{verdecchia2023systematic,10.1145/3639475.3640111}, we were able to find nine new green tactics and expanded the catalog of green architectural tactics, providing the frequency of adoption of these tactics and a code example to implement them. These findings provide a deeper understanding of how practitioners improve sustainability in software engineering for ML-enabled systems.  
This work is important for AI/ML researchers and software engineers who want to improve the sustainability of AI development and reduce environmental impact, which is currently a clear industrial need as the proliferation of companies and industry initiatives such as Green Coding \footnote{\url{https://www.green-coding.io/}} and the Green Software Foundation \footnote{\url{https://greensoftware.foundation/}} show, by helping clients, government agencies, and open-source communities to understand and optimize the carbon footprint of their digital services.

\smallskip
\textbf{Data availability statement.} For the sake of verifiability and replicability, we have made all the data and materials used in this study publicly available \cite{appendix}. The appendix includes the original dataset of \revised{205} ML projects, preprocessed data, and all scripts used for data collection, preparation, and analysis. Interested readers can use this material to assess the robustness/completeness of the process and encourage further research on adopting and updating green architectural tactics in ML-enabled systems.

\smallskip
\textbf{Structure of the paper.} Section~\ref{sec:related} reviews the related work. Section~\ref{sec:questions} describes the objectives of the study and introduces the research questions driving the study, while Section~\ref{sec:methodology} presents the methodology adopted to address them. Section~\ref{sec:result} reports the results obtained. Section~\ref{sec:discussion} discusses the implications of our findings. Finally, Section~\ref{sec:conclusion} concludes the paper and outlines directions for future work.

%% file: 2_RelatedWorks.tex
\section{Background and Related Work}
\label{sec:related}
In this section, we first introduce the concept of green architectural tactics, with particular attention to their definition, scope, and relationship to established architectural constructs. This description clarifies the role these tactics play in shaping environmentally sustainable design decisions within ML-enabled systems.

Second, we review the most relevant related work in sustainable software engineering and green AI, focusing on prior studies that have proposed, classified, or evaluated sustainability practices across the ML lifecycle. While existing literature has contributed significantly to the identification of green tactics and practices, we highlight a critical gap: the lack of empirical investigations into their adoption in real-world systems.

Our study addresses this gap by empirically assessing the frequency of green architectural tactics in real software projects, using LLM-based automation and human validation. In this way, we extend previous work by moving from the theoretical proposal of sustainability strategies to their practical application.

\subsection{\newchanges{Green Architectural Tactics}}
\newchanges{The increasing environmental footprint of ML-enabled systems has raised demand for systematic guidance on how to design such systems in a more energy-efficient and sustainable manner. In this context, the concept of green architectural tactics has gained traction as a lightweight and actionable means to capture design knowledge oriented toward environmental sustainability \cite{10.1145/3639475.3640111}. These tactics are intended to support architects and engineers in making informed design decisions that reduce energy consumption and, where possible, carbon emissions across the lifecycle of ML systems.}

\newchanges{Architectural tactics are high-level design decisions aimed at achieving specific quality attributes (QAs), such as performance, availability, or security \cite{bass2021software}. Unlike architectural patterns, which represent recurring structural solutions involving multiple design concerns \cite{gamma1995design}, tactics are focused, fine-grained strategies targeting a single QA. Patterns define broader architectural styles (e.g., client-server, layered), whereas tactics enable architects to shape specific system qualities by plugging into or modifying those patterns \cite{harrison2010how}.}

\newchanges{A green architectural tactic, as introduced by Järvenpää et al. \cite{10.1145/3639475.3640111}, is a tactic whose primary objective is to improve the environmental sustainability of ML-enabled systems, particularly by reducing energy consumption or carbon footprint. To ensure architectural relevance, a green tactic must influence system-level decisions, such as data flows, component interaction, or resource allocation. For instance, tactics like \textsl{`T1. Apply sampling techniques'} or \textsl{`T28. Use informed adaptation'} affect the amount of data processed or the way models are selected and executed, with direct implications for energy efficiency. Even tactics that seem implementation-level, such as \textsl{`T11. Use built-in library functions'}, are considered architectural when they guide significant design decisions impacting sustainability.}

\subsection{Related Work}
The interest in sustainable and green AI is increasing, and the SE and AI communities have published numerous articles to investigate through empirical studies which green practices reduce energy consumption and studies of the literature to catalog practices found in the literature in the entire ML pipeline \cite{verdecchia2023systematic,10.1145/3639475.3640111,natarajan2022theoretical,mittal2019survey}. However, most of the existing work in the literature focuses on investigating green practices, not assessing their adoption in real projects. 

One of the key factors in ML systems is the training time and size of such models, which can significantly impact computational costs and energy consumption \cite{schwartz2020green,castanyer2024design,shi2024greening}. Beyond these key factors, computational costs are also influenced by experiments, such as implementing models or algorithms \cite{hershcovich2022towards}, the type of library and framework \cite{georgiou2022green}, adjusting hyperparameters to assess the performance of a new model \cite{liao2022empirical}, containerization strategies \cite{hampau2022empirical} and the input type to make predictions \cite{haque2022ereba,chen2022nmtsloth}. In addition, there is little knowledge about how the carbon footprint of ML models is actually measured, reported, and evaluated \cite{castano2023exploring}. 

Salehi et al. \cite{salehi2023data}  conducted a systematic review to examine and classify data-centric approaches, highlighting practical applications and discussing the challenges and future directions of the green AI field. To capture what are the main existing approaches, Xu et al. \cite{xu2021survey} conducted a systematic review of Green Deep Learning approaches, focusing on reducing computational and energy costs in deep learning models while maintaining the same performance. Specifically, they mapped green AI approaches using a taxonomy of deep learning lifecycle stages and related artifacts to minimize the environmental impact of AI models. Following the work of Xu et al. \cite{xu2021survey}, Verdecchia et al. \cite{verdecchia2023systematic} conducted a systematic literature review expanding the domain not only to deep learning but to the entire domain of green AI, identifying a larger number of characteristics. More recently, Järvenpää et al. \cite{10.1145/3639475.3640111} surveyed the green practices proposed by Verdecchia et al. \cite{verdecchia2023systematic} so far, synthesizing and validating them through a focus group of three ML experts. As a result, they classified these tactics into clusters, expanded upon them, and identified additional tactics.

\novelty{As discussed in this section, much work has been done to advance knowledge on what approaches can improve the fields of sustainable and green SE.  However, to the best of our knowledge, our study is the first to investigate the actual adoption of green architectural tactics in real software projects, marking the primary scientific contribution of our work. We specifically study the set of tactics defined in the literature, developing and evaluating an LLM-based tool to (1) assess the adoption and frequency of these tactics in real-world projects and (2) expand the catalog of green AI tactics, providing reproducible code examples. Our research approach aims to deepen the understanding of sustainability, bridging the interdisciplinary gap between AI, software engineering, and sustainability, and building on previous research efforts.}


%% file: 3_Methodology.tex
\begin{figure*}[ht]
    \centering
    \includegraphics[width=1\linewidth]{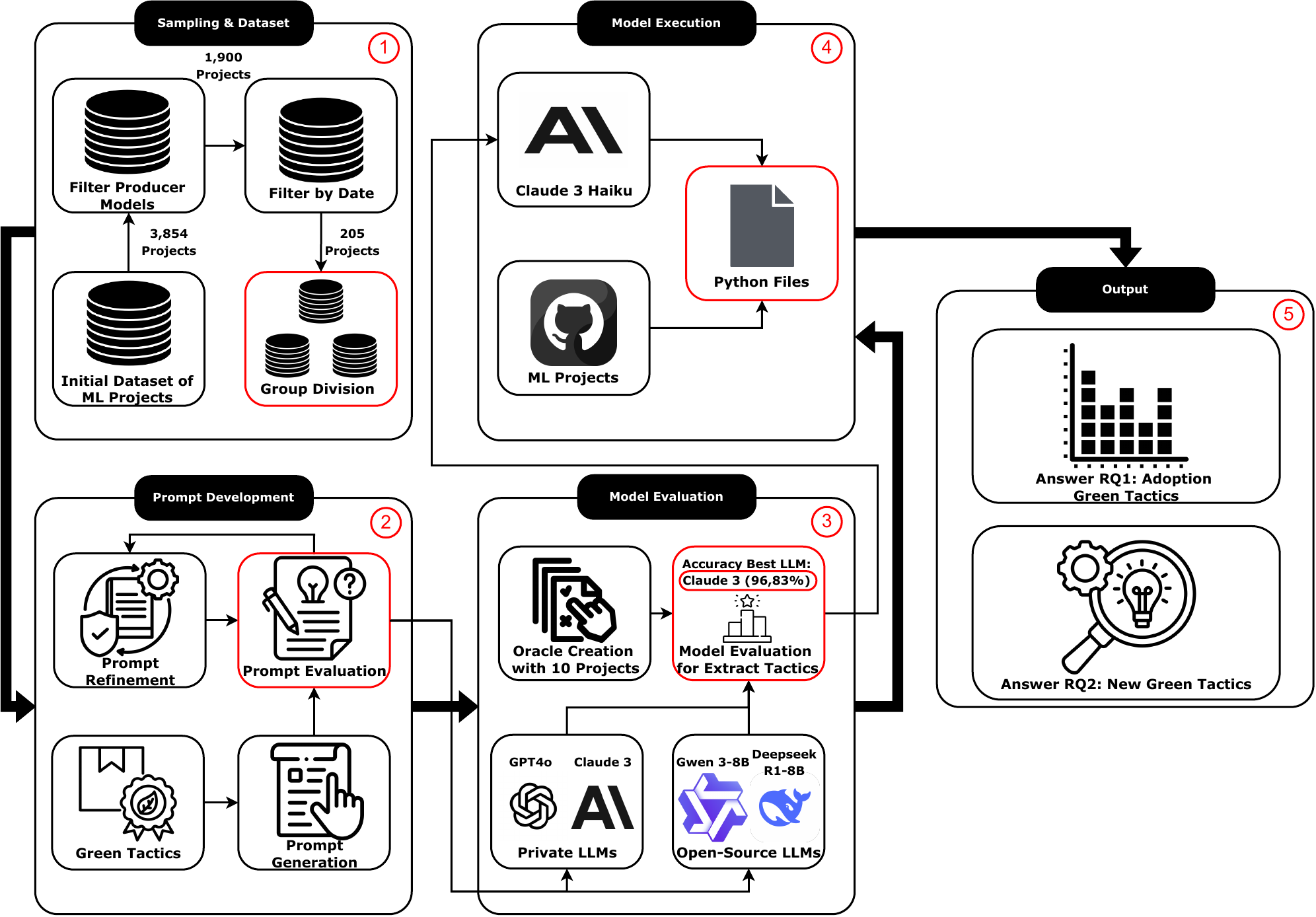}
    
    \caption{Overview of the research process.}
    \label{fig:overview}
\end{figure*}
\section{Research Goal and Questions}
\label{sec:questions}
To define our research goal, we follow the Goal-Question-Metric (GQM) approach~\cite{caldiera1994goal}. The goal of this study is to \emph{analyze} the adoption of green architectural tactics \emph{with respect to} their frequency and possible catalog extension, from the \emph{point of view} of researchers and practitioners, in the \emph{context} of ML-enabled system development.
More specifically, researchers seek to gain insights into the relevance and impact of green tactics in real software systems, which can inform future studies and guide the refinement of existing guidelines. For practitioners, understanding the most widely adopted tactics can help them implement sustainable practices in their projects, leading to reduced energy consumption, optimized system performance, and lower production costs. 

To achieve our research objective, we focus on two main research questions (\textbf{RQ}s). First, \textbf{RQ$_1$} seeks to determine the extent to which the green tactics for ML-enabled systems, as catalogued by Järvenpää et al. \cite{10.1145/3639475.3640111}, are adopted in the development of ML-enabled systems. Understanding this level of adoption is essential for bridging the gap between state-of-the-art research and the practices employed in real-world projects. Hence, we ask:

\sterqbox{RQ\textsubscript{1}. Adoption of Green Tactics in Practice}{To what extent are green architectural tactics for ML-enabled systems adopted in software projects?}

Beyond assessing the adoption of existing tactics that as been addressed in our previous work \cite{de2024developers}, \textbf{RQ$_2$} aims to expand the catalog by uncovering new tactics observed in software repositories. These insights will deepen understanding and provide researchers and practitioners with valuable knowledge, contributing to the development of best practices and guidelines for sustainable software. In particular, we ask:

\sterqbox{RQ\textsubscript{2}. Elicitation of Additional Green Tactics}{What additional green architectural tactics do developers adopt in building ML-enabled systems?}

To report the results of our exploratory mining study, we adhere to the standards for repository mining as outlined in the \emph{``Repository Mining''} and \emph{``General Standard''} categories of the \textsl{ACM/SIGSOFT Empirical Standards}.\footnote{\url{https://github.com/acmsigsoft/EmpiricalStandards}.}

\section{Research Method}
\label{sec:methodology}

This section describes our study with a quantitative (\textbf{RQ$_1$}) and qualitative (\textbf{RQ$_2$}) research approach to answer our \textbf{RQ}s. Figure \ref{fig:overview} overviews the main steps conducted in our study. 

\subsection{Sampling and Dataset Description}
\revised{We constructed our dataset by combining three previously published datasets of open-source ML-enabled systems. The first dataset, proposed by Gonzalez \etal~\cite{gonzalez2020state} and later revised by Rzig \etal~\cite{rzig2022characterizing}, contains 3,020 ML projects. The second dataset, \textsc{NICHE}~\cite{widyasari2023niche}, includes 572 ML projects. The third dataset include 262 projects~\cite{nahar2023product}.} Together, these datasets encompass a diverse array of ML projects, ranging from frameworks and tools to projects focused on creating ML models or integrating them into traditional software. However, not all projects within these datasets were suitable for our mining study. Specifically, our research aims to analyze sustainable tactics throughout the lifecycle of ML models. Therefore, we were required to filter out projects that do not involve the training of ML models, as they would not provide relevant insights for our study. To achieve this, we applied two main filters to the combined dataset of \revised{3,854 ML projects}, refining it to a subset that aligns with our research objectives.

\begin{itemize}[leftmargin=0.3cm]
    \item \textbf{Selection of projects that produce ML models.}
    We selected projects specifically focused on building and training ML models, e.g., those that enable the training of neural networks for tasks like face detection and pose estimation. We identified these projects by analyzing their APIs, looking for the presence of import statements for ML libraries and the invocation of related training methods \cite{mluniverse}. In this respect, we referred to a well-known and actively maintained list of AI \& ML libraries \cite{misiti2015awesome}, as previously done in literature \cite{gonzalez2020state}. After applying this criterion, our dataset was refined to include \revised{1,900 ML projects.}

    \smallskip
    \item \textbf{Selection of active projects with recent commits (2024).} The \revised{1,900} ML projects identified after the first filtering step were further refined based on their activity levels. We used the GitHub API \cite{GitHubREST} to identify projects that had been updated in 2024, following the methodology of previous studies \cite{bernardo2024machine}. This search, conducted in May 2024, narrowed our dataset to \revised{205 projects.}
\end{itemize}

\begin{figure}[h]
    \centering
    \includegraphics[width=1\linewidth]{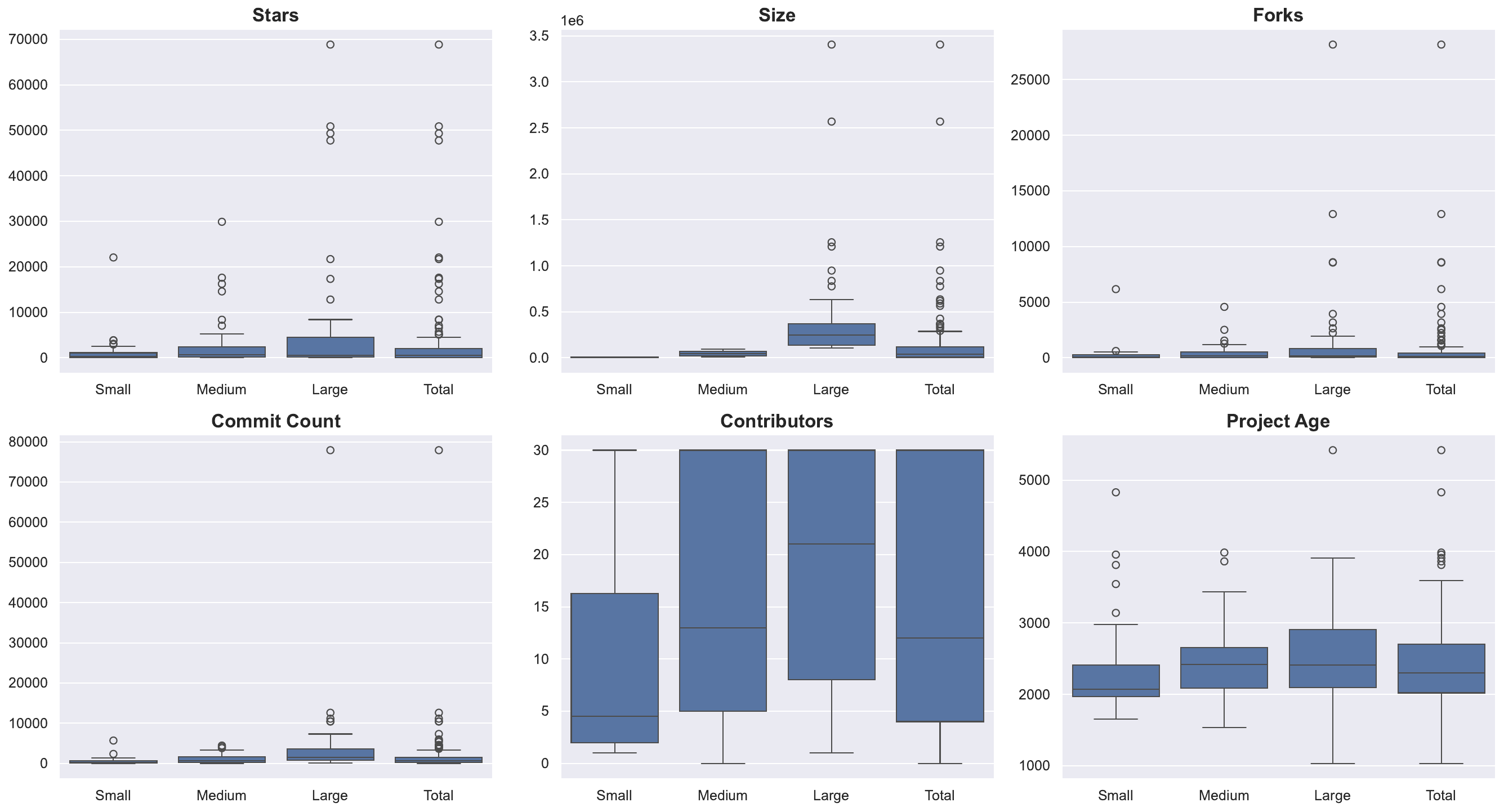}
    \caption{\revised{Distribution of Project Metrics Across Categories.}}
    \label{fig:characteristics}
\end{figure}

\revised{The final set comprises 205 ML projects. \Cref{table:project_metrics_info} summarizes the characteristics of these projects, including metrics such as age, number of stars, and number of contributors. Additionally, we use the lines of code (LOC) to classify projects into three size groups: small (LOC $\leq$ 10,000), medium (10,000 $<$ LOC $\leq$ 100,000), and large (LOC $>$ 100,000).
After this classification, we created four groups: \textsl{`Small'} with 67 projects, \textsl{`Medium'} with 72 projects, \textsl{`Large'} with 66 projects, and \textsl{`Total'} with 205 projects.
Considering the group \textsl{`Total'}, the average project age is 6.84 years, with a median of 6.45 years.
In terms of popularity, the projects have an average of 3,276.75 stars, with a median of 464 stars, indicating a wide range of visibility and adoption.
The average project size is 164,247.97 LOC, with a median of 41,205 LOC, reflecting significant variability in the scale of the projects.
On average, there are 18.97 contributors per project, with a median of 12 contributors.
The average number of forks is 678.71, with a median of 114, suggesting that while some projects are highly replicated, others have a more modest reach.
Finally, the average number of commits is 2,154.72, with a median of 691, highlighting varying levels of activity and development history among the projects.
To complement these aggregated statistics, Figure \ref{fig:characteristics} presents boxplots comparing each project metric across the three size categories (Small, Medium, Large) and the overall distribution (\textsl{Total}). As shown in the Figure \ref{fig:characteristics}, each metric exhibits high variance within and across categories, illustrating the heterogeneity of the analyzed ML repositories. For example, the boxplots for \textsl{Stars}, \textsl{Forks}, and \textsl{Commit Count} (top row of the figure) reveal positively skewed distributions with several high-value outliers, while \textsl{Contributors} and \textsl{Project Age} (bottom row) display narrower spreads and more compact central tendencies. The selected projects show diverse metrics, which is satisfactory for the study's goals. This allows generalizable conclusions about the presence of green architectural tactics in different types of ML projects.}

\begin{table}[h]
\caption{\revised{Average (AVG) and Median (Med) of each group of projects.}}
\label{table:project_metrics_info}
\centering
\footnotesize
\begin{tabular}{|l|c|c|c|c|}
\rowcolor{black}
\textcolor{white}{\textbf{Metric}} & \textcolor{white}{\textbf{Small}} & \textcolor{white}{\textbf{Medium}} & \textcolor{white}{\textbf{Large}} & \textcolor{white}{\textbf{Total}} \\
\hline
\#Projects & 67 & 72 & 66 & 205 \\
\hline
\rowcolor{gray10} Age (Avg) & 6.32 & 6.94 & 7.26 & 6.84 \\
\hline
Age (Med) & 5.7 & 6.67 & 6.94 & 6.45 \\
\hline
\rowcolor{gray10} Stars (Avg) & 1,004.9 & 2,286.93 & 6,662.82 & 3,276.75 \\
\hline
Stars (Med) & 245 & 581.5 & 563.5 & 464\\
\hline
\rowcolor{gray10} Size (Avg) & 3,568.28 & 46,879.25 & 455,400.8 & 164,247.97 \\
\hline
Size (Med) & 2,779.0 & 41,825.5 & 254,595.5 & 41,205 \\
\hline
\rowcolor{gray10} Contributors (Avg) & 8.81 & 16.69 & 31.77 & 18.97 \\
\hline
Contributors (Med) & 5 & 13 & 21 & 12 \\
\hline
\rowcolor{gray10} Forks (Avg) & 224.16 & 367.56 & 1,479.59 & 678.71 \\
\hline
Forks (Med) & 62 & 132 & 157 & 114 \\
\hline
\rowcolor{gray10} Commits (Avg) & 492.43 & 1204.22 & 4,879.11 & 2,154.72 \\
\hline
Commits (Med) & 246 & 744.5 & 1,528.5 & 691 \\
\hline
\end{tabular}
\end{table}

\subsection{Selection of Architectural Green Tactics} \label{sec:green_tactics}
After selecting the ML projects, we proceeded to identify the architectural tactics that would be the focus of our analysis. We chose to examine the green architectural tactics for ML-enabled systems defined by Järvenpää et al. ~\cite{10.1145/3639475.3640111}, which build upon the foundational work of Verdecchia \etal~\cite{verdecchia2023systematic}. This selection was based on the fact that the catalog provided by Järvenpää et al. \cite{10.1145/3639475.3640111} is, to our knowledge, the most comprehensive resource available, encompassing green practices across the entire ML pipeline. While other studies have addressed domain-specific aspects or isolated strategies \cite{10.1145/3639475.3640111,natarajan2022theoretical,mittal2019survey}, the catalog by Järvenpää et al. \cite{10.1145/3639475.3640111} organizes 30 green architectural tactics into six clusters, each targeting different aspects of energy efficiency and sustainability within an ML pipeline. These clusters focus on various aspects of energy efficiency and sustainability of a pipeline of ML-enabled systems.

From these candidate tactics, our study focuses on those pertaining to three specific clusters: \textsl{`Algorithm design'}, \textsl{`Model optimization'}, and \textsl{`Model training'}, encompassing a total of 15 tactics. \revised{These tactics were not selected based on perceived importance, but based on their \emph{detectability through static analysis of source code}. Because our goal is to investigate the feasibility of automated, code-centric identification of green architectural practices, we only included tactics whose implementation leaves \emph{explicit and unambiguous code-level footprints} that can be observed directly in GitHub repositories. Several tactics in the original catalog cannot be analyzed through code alone.}

\revised{Many data-centric or system-level practices depend on artifacts or processes that are \emph{not encoded in the source code} and therefore cannot be reliably inferred. For example, a tactic like \textsl{`Remove redundant data'} requires access to the original dataset, intermediate versions, or preprocessing logs to determine whether redundancy has actually been removed, information that is typically not present in repositories. Similarly, the practice of using \textsl{`Federated learning'} requires deployment topology, client–server training orchestration, device-local data constraints, and aggregation protocols, which reside in configuration files, infrastructure definitions, or runtime metadata rather than in the model code itself. Conversely, the tactics we selected are those whose implementation naturally appears in the code. For example, \textsl{`Clean and Standardize Data'} is implemented through explicit preprocessing APIs (e.g., scaling, normalization, type casting, handling missing values), which appear directly in the code and can be consistently detected. On a similar note, \textsl{`Employ Parallel Training'} relies on recognizable code constructs such as \texttt{torch.nn.DataParallel}, \texttt{torch.distributed}, or multiprocessing routines that are observable at the code level without external information. Thus, the boundary of our analysis is determined by the extent to which a tactic's implementation is \emph{visible and verifiable in source code}. Practices requiring dataset inspection, infrastructure configuration, hardware specifications, runtime profiling, or organizational decisions fall outside the scope of automated static analysis and were therefore excluded. At the same time, we would also acknowledge that some selected tactics, e.g. \textsl{`Clean and Standardize Data'}, \textsl{`Consider Use of Data Augmentation'}, or \textsl{`Use Early Stopping'}, are also common ML best practices aimed at improving robustness and model performance. Their inclusion in our study does not stem from a claim that these practices are inherently green, but from the fact that they are formally classified as green architectural tactics in the catalog by Järvenpää et al.~\cite{10.1145/3639475.3640111}. In that taxonomy, these practices are considered to have sustainability benefits because they can reduce computational waste, stabilize training dynamics, or lower the need for repeated or resource-intensive model training. As an example, \textsl{`Use Early Stopping'} helps avoid unnecessary training epochs once performance plateaus, thereby reducing computational load and energy consumption. Our goal is therefore not to establish the greenness of these tactics, but to systematically analyze the subset of tactics defined in the existing literature that are detectable through source code.}


\begin{table}[htbp]
    \caption{\newchanges{Green architectural tactics for ML-Enabled Systems selected in the scope of our work and the Large Language Model (LLM) detection rationale.}}
    \centering
    \resizebox{\textwidth}{!}{
    \begin{tabular}{|>{\raggedright}p{3.5cm}|>{\raggedright}p{4.5cm}|>{\raggedright\arraybackslash}p{5.5cm}|}
        \rowcolor{black}
        \textcolor{white}{\textbf{Tactic}} & \textcolor{white}{\textbf{Description}} & \textcolor{white}{\textbf{LLM Detection Rationale}} \\\hline
        T6. Choose an energy-efficient algorithm & Choose the most energy-efficient algorithm that achieves sufficient level of accuracy. & LLMs can recognize mentions of algorithm types and associate them with known energy profiles. \\\hline
        \rowcolor{gray10} T7. Choose a lightweight algorithm alternative & If possible, choose lighter alternatives of existing algorithms. & LLMs are able to infer that 'lightweight' implies lower computational load, especially in model selection contexts. \\\hline
        T8. Decrease model complexity & Decrease the complexity of an ML model. & LLMs detect patterns indicating efforts to simplify model architecture (e.g., fewer layers, simpler structures). \\\hline
        \rowcolor{gray10} T9. Consider reinforcement learning for energy efficiency & Use reinforcement learning to optimize energy efficiency at run time. & LLMs understand the concept of 'optimization' and can link RL applications to dynamic energy management strategies. \\\hline
        T10. Use dynamic parameter adaptation & Design parameters that are dynamically adapted based on the input data. & LLMs can detect descriptions of hyperparameter tuning that adapt based on input data, which aligns with energy efficiency goals. \\\hline
        \rowcolor{gray10} T11. Use built-in library functions & Use built-in libraries for ML models if possible. & LLMs can identify common APIs or functions from well-known libraries (e.g., NumPy, PyTorch) and flag when they're used. \\\hline
        T12. Set energy consumption as a model constraint & Consider energy consumption as one predetermined parameter for optimizing the ML model. & LLMs interpret explicit mentions of constraints during training as optimization objectives, including energy constraints. \\\hline
        \rowcolor{gray10} T13. Consider graph substitution & Replace energy-intensive model parts with similar, but less energy-consuming parts. & LLMs detect substitutions in model components that imply reduced resource usage, especially if semantically equivalent. \\\hline
        T14. Enhance model sparsity & Reduce the number of model parameters or set their values to zero. & LLMs recognize sparsity-related terminology and reduction in parameter counts, linking them to lower computation. \\\hline
        \rowcolor{gray10} T15. Consider energy aware pruning & Prune neural networks starting from the most energy-intensive layer. & LLMs detect techniques such as 'pruning' and associate layer-level energy metrics with selective removal. \\\hline
        T16. Consider transfer learning & Use pre-trained ML models for other similar tasks. & LLMs can infer reuse from pretraining references and match it to the concept of transfer learning. \\\hline
        \rowcolor{gray10} T17. Consider knowledge distillation & Use knowledge from a large ML model to train a smaller model. & LLMs identify 'teacher-student' language or simplification processes as indicators of knowledge distillation. \\\hline
        T18. Use quantization-aware training & Convert high-precision data types to lower precision during training. & LLMs recognize numeric precision reductions during training as quantization, which lowers computational demand. \\\hline
        \rowcolor{gray10} T19. Use checkpoints during training & Use checkpoints to avoid a knowledge loss in case of a premature termination. & LLMs detect checkpointing mechanisms and relate them to training resumption or fault tolerance. \\\hline
        T20. Design for memory constraints & Consider possible memory constraints during training. & LLMs infer memory constraints from language indicating hardware limits or optimization strategies. \\\hline
    \end{tabular}
    }
    \label{table:tactics}
\end{table}

\subsection{Extraction of Architectural Green Tactics}
\label{sec:extractionTool}
\newchanges{Detecting the presence of green architectural tactics in software repositories poses significant challenges, as it requires discerning whether developers have implemented specific practices to support sustainability efforts. Developers rarely annotate code with explicit sustainability intentions, making it difficult to discern whether a particular implementation reflects a green tactic.}
\newchanges{In exploring alternatives, we considered both manual and automated approaches. Manual inspection is largely impractical for our context because it is time-consuming, prone to errors, and requires a deep familiarity with diverse code languages. Additionally, recognizing a tactic may involve interpreting detailed code instructions or entire blocks, which can introduce ambiguity and inconsistency.
Automated analysis, while more scalable, presents its own challenges. For example, let consider the case of the design pattern detection \cite{mayvan2017state}, where automated tools and algorithms are used to identify recurring structures in the code that correspond to known design patterns. This task is non-trivial because the same pattern can be implemented in multiple syntactic forms, and its recognition often requires an understanding of object-oriented principles and semantic relationships between classes and methods. Similarly, the detection of green architectural tactics involves identifying higher-level design decisions that are not explicitly documented and may be scattered across different parts of the codebase. 
As such, the task of identifying green architectural tactics can be framed as a \textbf{complex pattern recognition problem}, for which ML is a natural fit. Among the available ML techniques, we adopted LLMs due to their ability to understand context, infer semantics across non-local code elements, and generalize patterns from natural language definitions to real code structures. LLMs offer a promising basis for detecting design intent at the tactical level that is not explicitly marked but is manifested in the structure and behavior of the code.}


\minor{More specifically, we mined the latest snapshot of the repositories associated with the selected software projects, extracting all \textsl{Python} files, i.e., files that may contain source code related to ML model creation, training, and deployment. To detect architectural green tactics, we applied LLMs at the file level, processing one \textsl{Python} file per invocation and providing the complete file content (UTF-8) concatenated to a fixed analysis prompt. This approach processes each file in a single invocation and aims to preserve intra-file dependencies (e.g., imports, class and function definitions, and comments), assuming that file sizes fall within the context window of the models being evaluated. However, during execution, we did not record token-level statistics in the log file, and alternative strategies, such as chunking or hierarchical processing, could affect the information available to the model and potentially lead to different classification outcomes. Future studies could investigate the impact of these techniques on classification results.}


LLMs are highly effective at interpreting source code, enabling them to identify subtle patterns and context-specific practices within these files \cite{lopes2024commit}. In particular, the advanced architecture and extensive training of LLMs allow them to capture complex relationships and subtle patterns within code, making them potentially effective at identifying context-specific instances of green tactics in ML projects. Indeed, by understanding the context of the source code implemented within a file, e.g., recognizing when an algorithm is optimized for energy efficiency or when a model is being pruned to reduce computational load, LLMs may accurately detect the application of green tactics even when these are not explicitly documented. Several factors influenced our decision to use LLMs over other ML models. First, LLMs are effective in natural language processing tasks due to their architecture and training on diverse language datasets, which enhances their ability to identify and interpret code within files \cite{hou2023large}. Additionally, LLMs are adept at processing and analyzing large volumes of code, making them well-suited for the scale of our study. Furthermore, LLMs can execute and solve tasks based on instructions without requiring additional training data \cite{zhang2023prompting}, which was a critical requirement in our case, given the absence of a labeled dataset specifically tailored for detecting green architectural tactics in software repositories. To this aim, we devised an LLM-based mechanism aiming at detecting the application of green architectural tactics within software code. Our approach employs prompt engineering, where carefully crafted input prompts guide the LLM to generate outputs that accurately reflect the presence of tactics \cite{zhou2022large,radford2019language}. 


Based on the objectives of our work, we developed a prompt to instruct the LLM in achieving two primary objectives: (1) \textsl{`Identify Existing Architectural Green Tactics'}, the LLM was directed to identify the 15 sustainable tactics listed in Table \ref{table:tactics} and determine their presence in the code; (2) \textsl{`Discover Additional Architectural Green Tactics'}, the LLM was tasked with detecting any sustainable tactics not included in Table \ref{table:tactics}. \newchanges{Finally, for each tactic in Table \ref{table:tactics}, we described the rationale of LLMs to detect tactics with a version of the code. To conduct the entire prompt development, we followed the recommendations of the PRIMES framework \cite{de2024framework}.}

\begin{figure}[h]
    \centering
    \includegraphics[width=0.9\linewidth]{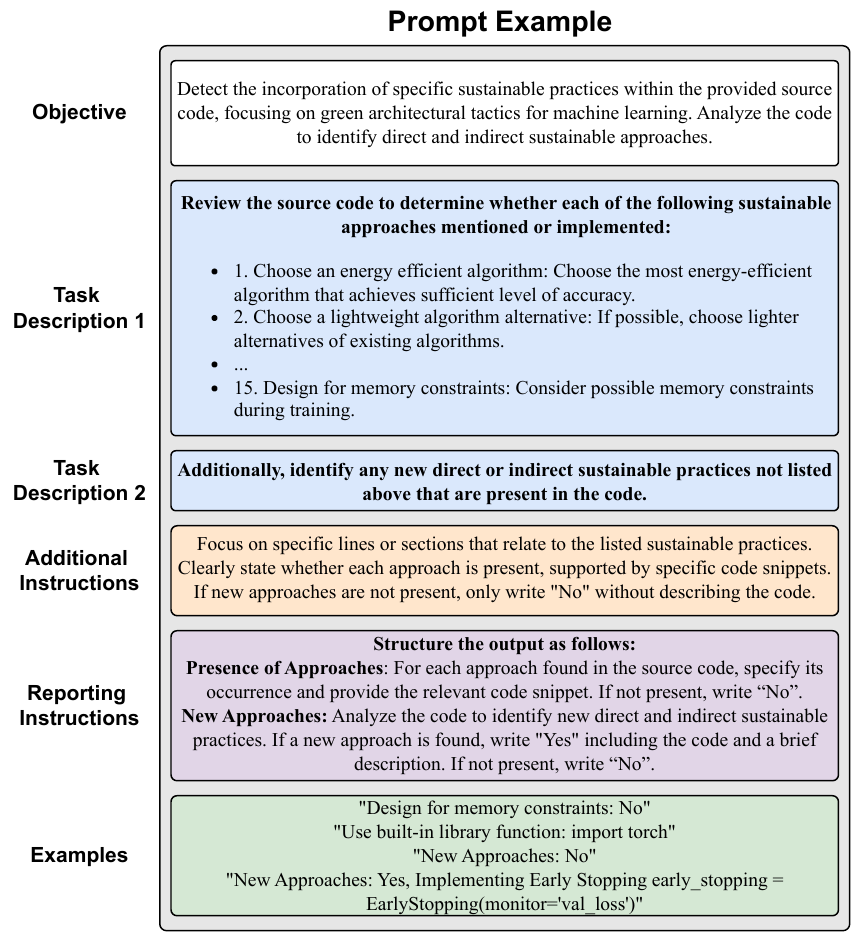}
    \caption{The prompt designed to extract architecture green tactics from software repositories.}
    \label{fig:prompt}
\end{figure}

To accomplish these objectives, we structured the prompt into the following five key components, as illustrated also in Figure \ref{fig:prompt}:

\begin{description}[leftmargin=0.3cm]
    \item[\textsl{`Objective'}.] The first part of the prompt defines the overarching goal, instructing the LLM to examine the source code to detect both direct and indirect sustainable practices implemented within the code changes analyzed, with a particular emphasis on green architectural tactics for machine learning;

    \smallskip
    \item[\textsl{`Task Description'}.] We outlined two specific tasks for the LLM. The first task requires the LLM to detect the 15 sustainable tactics from Table \ref{table:tactics} and verify their presence in the code. The second task is to uncover new sustainable practices not previously documented.

    \smallskip
    \item[\textsl{`Additional Instructions'}.] This part details information on how the LLM should analyze the code and generate the final output;

    \smallskip
    \item[\textsl{`Reporting Instructions'}.] The guidelines instruct the LLM to clearly present its findings, indicating the presence of each tactic and including relevant code snippets. 
    If a tactic is not detected, the LLM should explicitly indicate its absence by associating a \emph{``No''} with the tactic in the output;

    \smallskip
    \item[\textsl{`Examples'}.] The final part provides examples of how to correctly format and report the results. The examples serve as templates, guiding the LLM in generating output consistent with the desired structure. An example is provided for each possible scenario.
\end{description}

\revised{We experimented with the prompt using four LLMs, two proprietary models, GPT-4o from OpenAI~\cite{gptkey} and Claude~3 Haiku from Anthropic~\cite{haikukey}, and two open-source models, Qwen3-8B and DeepSeek-R1-8B, to determine the most suitable model for our analysis. These models were selected for their recognized accuracy in task resolution and their ability to effectively interpret complex contexts, such as source code~\cite{chang2024survey}. The two proprietary LLMs were executed via their respective APIs, whereas the open-source models were deployed locally on a dedicated server equipped with an NVIDIA RTX 4090 GPU (24~GB VRAM), an AMD Ryzen~9~7950X CPU with 16 cores and 32 threads, and 32~GB of DDR5 RAM. All model inferences were executed using their default configurations, without any modification of temperature, token limits, or sampling parameters, to ensure comparability and reproducibility. Detailed information about the models employed, including their type, provider, and configuration, is reported in Table~\ref{table:llm_specs}. The development of the prompt was an iterative process involving multiple refinements to enhance its accuracy and effectiveness. The first two authors of the paper conducted this process through collaborative sessions, where initial versions of the prompt were tested on a subset of projects. Feedback from these tests informed adjustments to the instructions and examples provided to the LLM. After nine iterative refinements, the prompt evolved from a concise initial version into a more detailed formulation that included clearer task instructions, explicit definitions of architectural tactics, and illustrative examples. Although this resulted in a longer prompt, the additional context significantly improved classification accuracy and consistency across repositories. The evolution of the prompt, including its various iterations, is documented in the appendix~\cite{appendix}.}


\begin{table}[h]
\caption{\revised{Large Language Models Characteristics all with Default Parameters.}}
\label{table:llm_specs}
\centering
\footnotesize
\rowcolors{1}{gray!15}{white}
\begin{tabular}{|c|c|c|c|c|}
\rowcolor{black}
\textcolor{white}{\textbf{Characteristic}} &
\textcolor{white}{\textbf{Qwen3-8B}} &
\textcolor{white}{\textbf{Deepseek-R1-8B}} &
\textcolor{white}{\textbf{GPT4o}} &
\textcolor{white}{\textbf{Claude 3 Haiku}} \\
\hline
Organization & Qwen & Deepseek & Open-AI & Anthropic \\
\hline
Params & 8.2B & 8.19B & Unknown & Unknown \\
\hline
Release & April 2025 & April 2025 & May 2024 & March 2024 \\
\hline
Temperature & 0.6 & 0.6 & 1.0 & 1.0 \\
\hline
Context Length & 33k & 131k & 128k & 200k \\
\hline
I/O 1M Tokens Cost & None & None & \$5/15 & \$0.25/1.25 \\
\hline
\end{tabular}
\end{table}

As a final note, we designed the green architectural tactics mechanism with flexibility in mind, separating the code into two distinct components: one that handles project analysis and another that interfaces with the LLMs. This modular approach allows for easy integration of newer, more accurate, and higher-performing LLMs in the future. As such, the mechanism we developed can be further extended to support future work in the field.

\subsection{Model Validation}
\label{sec:modelValidation}
\revised{To evaluate the suitability of using an LLM to extract architectural green practices, we conducted an oracle-based validation.} This process involved testing the prompt with both GPT-4o and Claude 3 Haiku across a representative subset of projects. We specifically applied random sampling to ensure unbiased selection~\cite{baltes2022sampling,thompson2012sampling}, identifying \revised{21} projects from our initial dataset, which included \revised{141} Python files. \revised{To ensure a reliable oracle, two authors independently classified tactics within a subset of 10 projects, which represents approximately 10\% of the total set.} By analyzing the selected files, the authors manually constructed an oracle by classifying the 10 projects according to the presence or absence of sustainable tactics. 
This process involved examining the code to identify specific implementations of the green tactics within the scope of our work. Leveraging their research experience in sustainable software engineering, the authors compared each identified practice with those cataloged by Järvenpää et al. \cite{10.1145/3639475.3640111}. In doing so, they used the exemplary code snippets provided in the catalog as a reference to determine whether a given piece of code implemented a sustainable tactic. It is important to note that this preliminary evaluation focused exclusively on practices already cataloged in the literature, as these could be reliably identified and manually classified using the well-defined criteria and examples provided by prior research \cite{10.1145/3639475.3640111}. In contrast, the identification and validation of previously undocumented tactics lack established definitions and examples, making it difficult to manually classify them with the same level of confidence. Consequently, we deferred the classification of new tactics to a subsequent phase of data analysis, which was guided by the insights provided by the LLM (see Section \ref{sec:extraction}). Indeed, the prompt design not only allowed to identify potential new tactics but also to provide detailed explanations and examples that justified why these practices might qualify as sustainable. This approach allowed for a more informed and rigorous evaluation process in the later stages of the study. 

\minor{This approach is standard in qualitative research, where employing a subset of 10\% for inter-rater reliability is common practice to enhance the reliability of the coding scheme \cite{o2020intercoder}. The initial human checks indicated substantial agreement between the two authors. To quantify agreement, we computed Cohen’s kappa~\cite{cohen1960coefficient} on the dual-coded subset, using individual (\textit{file}, \textit{tactic}) decisions as the unit of analysis and binary labels indicating the presence or absence of a tactic. Initial independent annotations yielded a Cohen’s kappa of 0.44, indicating moderate agreement. Any disagreements were addressed through discussions held across three meetings, thereby establishing a clear ground truth. These discussions allowed us to create more explicit examples for each tactic, clarifying distinctions between closely related tactics (e.g., T7: Choose a lightweight algorithm alternative vs. T20: Design for memory constraints) and linking definitions to specific tactics (e.g., T10: Use dynamic parameter adaptation, which might be confused with other parameter adaptations). After this consensus phase, full agreement was reached on the subset (Cohen’s kappa = 1.00). The double-blind classification process and the standardized consensus for classification, despite potential biases, served as effective measures to mitigate subjective bias and to enable continued classification by a single author. Then, the first author undertook the classification of all 141 files, a task that required over 150 hours of dedicated work.}


To better illustrate the manual classification process, consider a Python file that implements a neural network model's classification capabilities quantitatively but also provides computational load, which aligns with the tactic of \textsl{`Consider energy-aware pruning'}. The author would then classify the file as containing this sustainable tactic. Conversely, if the code lacks such optimizations and does not exhibit any tactics from the catalog, it would not be classified as sustainable tactics. 

\revised{Upon construction of the oracle, we executed GPT-4o, Claude 3 Haiku, Qwen3-8B, and DeepSeek-R1-8B  with the prompt described in Section \ref{sec:extractionTool}, collecting their responses without modifying the temperature or any other LLM parameters. \minor{Overall classification accuracy was computed at the level of individual (\textit{file}, \textit{tactic}) pairs. For each analyzed Python file, the LLM produced a binary decision (Yes/No) for each of the 15 predefined green architectural tactics (T6-T20). A prediction was considered correct if the model’s label matched the oracle label for the same tactic in the given file. Accuracy was calculated as the ratio between the number of correct predictions and the total number of evaluated file--tactic decisions across the validation set, corresponding to a micro-averaged accuracy.} Through this procedure, we could iteratively refine the prompt and evaluate its impact on the results. Additionally, using the prompt defined in Section \ref{sec:extractionTool} not only enabled us to assess the classification capabilities of the models from a quantitative perspective but also offered insights into their performance in identifying additional green architectural tactics. Although this was not the primary focus of our evaluation, observing how each model handled the identification of new tactics provided valuable trends and considerations that informed our selection of the most suitable model for our study. As a result of this evaluation, the four models demonstrated high accuracy, each with its own distinct strengths.}

\begin{table}[h]
\caption{\revised{Classification Accuracy of Evaluated Large Language Models.}}
\label{table:llm_accuracy}
\centering
\footnotesize
\rowcolors{1}{gray!15}{white}
\begin{tabular}{|c|c|c|}
\rowcolor{black}
\textcolor{white}{\textbf{Model}} &
\textcolor{white}{\textbf{Type}} &
\textcolor{white}{\textbf{Accuracy (\%)}} \\
\hline
DeepSeek-R1-8B & Open-source & 90.48 \\
\hline
Qwen3-8B & Open-source & 91.26 \\
\hline
GPT-4o & Proprietary & 93.66 \\
\hline
Claude 3 Haiku & Proprietary & \textbf{96.83} \\
\hline
\end{tabular}
\end{table}

\revised{The results of the extended evaluation, including open-source LLMs, are reported in Table~\ref{table:llm_accuracy}. Claude 3 Haiku achieved the highest accuracy (96.83\%), followed by GPT-4o (93.66\%), Qwen3-8B (91.26\%), and DeepSeek-R1-8B (90.48\%). These findings confirm that modern open-source models can achieve accuracy comparable to proprietary models, although with notable differences in interpretability and discovery capabilities. In the oracle-based comparison, misclassifications for Claude 3 Haiku were rare and primarily occurred in borderline cases, where the implementation of a tactic was implicit or could reasonably be mapped to more than one tactic definition (e.g., when code patterns overlapped between T7: \textit{Choose a lightweight algorithm alternative} and T20: \textit{Design for memory constraints}). Importantly, we did not observe hallucinated tactics, i.e., predictions outside the predefined catalog described in Section \ref{sec:green_tactics}. This indicates that residual errors are mainly due to semantic ambiguity in the code or conceptual overlap between tactics, rather than to spurious or fabricated classifications. GPT-4o demonstrated strong quantitative performance but provided primarily analytical outputs with limited explanatory depth. When identifying a tactic, GPT-4o typically returned the relevant code snippet but offered minimal reasoning, and it did not identify any previously undocumented sustainable practices. This limited its suitability for interpretability-oriented analysis. Claude 3 Haiku, in contrast, not only achieved the highest accuracy but also produced substantially richer natural-language explanations that clarified the rationale behind each classification. Although it did not always isolate exact code snippets, opting instead for descriptive references, it demonstrated a stronger ability to identify additional potential sustainable tactics beyond those already documented. Occasional repetitions of known tactics were observed, yet the overall quality and clarity of Haiku's explanations made it the most effective model for supporting both classification and exploratory analysis.}

\revised{Open-source models (Qwen3-8B and DeepSeek-R1-8B) respectively performed on the classification task but showed limitations in fine-grained code reasoning and explanatory coherence. Their outputs tended to be shorter and less consistent, and their ability to identify potential new tactics was more limited than that of proprietary alternatives.}

Based on these findings, we selected Claude 3 Haiku for its higher accuracy, its ability to identify additional sustainable tactics, and its lower cost. Although Haiku did not always return the specific code associated with the tactic, we implemented a mechanism that allowed us to track the repository and file in which the tactic was detected. This ensures that, even if the code snippet is not provided explicitly, the tactic could still be located within the source file. \newchanges{As a final remark, it is worth reporting that, while we have opted to focus on a single LLM for consistency and cost-effectiveness, future research could investigate ensemble methods that exploits the strengths of multiple LLMs to further enhance the detection and explanation of green tactics.}

\subsection{Data Analysis}
\label{sec:extraction}
To facilitate data analysis, we developed a Python script ~\cite{appendix} that systematically traverses the files in each repository and records key information. For each project, it generates a \texttt{csv} file, with each row corresponding to an analyzed file. The \texttt{csv} file captures (1) the file name and (2) the output generated by Claude 3 Haiku. 

To address \textbf{RQ$_1$}, the implemented script processed Haiku's responses by segmenting the first part of the output, specifically the classification of existing tactics, into separate columns, with each column representing one of the tactics considered in the study. After collecting all responses, the data was aggregated into a single file that tracked the occurrence of each tactic. The script calculated the total frequency of each tactic across all projects and used this information to address the first \textbf{RQ}, providing an overview of how green architectural tactics are adopted in the ML-enabled systems considered. \revised{Beyond descriptive frequency analysis, we further investigated whether the adoption of each tactic is statistically associated with project-level characteristics by fitting logistic regression models, a standard approach for modeling binary adoption outcomes \cite{hosmer2013applied}.} \revised{Because \textbf{RQ$_1$} concerns the detection of predefined tactics from an established catalog, the task constitutes a closed-set classification problem. As such, the LLM cannot introduce novel or unsupported tactics: any tactic not recognized is automatically labeled as ``No’’. This structure inherently limits hallucination risks, as confirmed by the oracle-based evaluation conducted in Section~\ref{sec:modelValidation}, where multiple LLMs and prompt versions were compared against a manually constructed ground truth. Claude 3 Haiku achieved very high accuracy in identifying known tactics, providing a reliable basis for the large-scale analysis. For this reason, an additional manual validation step was not required, as hallucination risks had already been quantified and controlled during the oracle-based model validation phase.}

To address \textbf{RQ$_2$}, we processed Haiku's responses to isolate the output provided for the second task, specifically, the identification of additional tactics not covered in the original catalog. \newchanges{Given the non-deterministic nature of LLMs, which can give rise to hallucinatory or spurious suggestions, we applied a manual validation process to mitigate this risk. Each candidate's tactic was examined to ensure it reflected a consistent, actionable sustainability practice. The manual analysis excluded responses containing hallucinations and improved the accuracy and reliability of the tactics presented in the remainder of the paper.} We first filtered out any responses where the model did not suggest new tactics, focusing only on instances where the model provided potential new sustainable practices. This could be done automatically, as the prompt required the model to answer \emph{``No''} when no additional tactics were found. From this filtering step, we obtained \revised{1,970 responses}, which were analyzed through iterative content analysis sessions \cite{krippendorff2018content}. More specifically, the first author manually reviewed each response to determine whether the identified tactics were novel or pre-existing. Each tactic was initially assigned one of three possible grades: \emph{`No'}, \emph{`Maybe'}, or \emph{`Yes'}. Tactics classified as \emph{`No'} were those already included in the existing catalog, unrelated to ML systems, or not considered sustainable. Tactics classified as \emph{`Maybe'} refer to those with ambiguities due to insufficiently detailed descriptions that make it difficult to assess their applicability. These tactics may also address potential sustainability factors, such as energy efficiency, that necessitate further interpretation or contextual understanding and would benefit from a discussion among the authors. Finally, tactics classified as \emph{`Yes'} were clearly described and aligned with sustainable practices for ML-enabled systems.

This first content analysis session was supervised by the second author of the article, who took an active role in guiding the process, providing oversight and ensuring that the classification criteria were consistently applied. Together, the first and second authors also discussed the suitability of the tactics labeled \emph{`Maybe'}, deliberating on whether to reclassify or exclude them based on their potential contribution to sustainable practices. Following this initial session, a subsequent content analysis iteration was conducted to further refine the classification. This iteration involved re-evaluating the \emph{`Maybe'} tactics in light of additional insights gained during the discussions. The authors aimed to reach a consensus on all tactics, carefully discussing any disagreements to ensure that only practices clearly aligned with sustainable principles were included in the final set. \revised{In the manual inspection process, we observed three main types of issues: (i) overly generic recommendations that did not constitute concrete architectural tactics, (ii) suggestions that rephrased or duplicated tactics already present in the existing catalog, and (iii) practices that were not clearly related to environmental sustainability. These cases were systematically excluded, leaving only well-defined, sustainability-relevant, novel tactics in the final catalog.}

\revised{From Claude 3 Haiku’s responses, we initially obtained 1,970 candidate tactic instances, where each instance corresponds to a single tactic suggested by the model in response to the analysis of a specific source-code file. At this stage, candidates were not deduplicated, and the same tactic could therefore appear multiple times across different files or projects. Of these 1,970 instances, 1,416 were classified as \emph{`No'} because they either did not represent sustainable practices, were not applicable to ML-enabled systems, or duplicated tactics already present in the existing catalog. Notably, 119 of the discarded instances corresponded to tactics already included in the original catalog. The remaining instances were classified as \emph{`Yes'} (558) and \emph{`Maybe'} (151). Following a consensus-based discussion among the authors, all \emph{`Maybe'} instances were excluded due to insufficient detail or ambiguity. This process yielded 558 validated tactic instances, which were subsequently subjected to in-depth qualitative analysis and consolidated into a smaller set of distinct green architectural tactics.}
\revised{To determine whether these tactics were truly green and to identify novel contributions, we applied a structured multi-step validation process. For each candidate tactic we: (1) assigned a concise and representative title; (2) mapped the tactic to one of the sustainability groups defined by Järvenpää et al.~\cite{10.1145/3639475.3640111}; (3) wrote a detailed description capturing its intent, mechanism, and sustainability rationale; (4) documented the tactic in a shared repository containing all validated items; (5) extracted relevant code examples illustrating how the tactic was implemented in practice; (6) performed literature triangulation by searching for conceptual or empirical counterparts in prior work on software sustainability, energy-efficient ML, green AI, and environmentally aware software engineering; and (7) assigned quality attributes informed by definitions from Järvenpää et al.~\cite{10.1145/3639475.3640111}. The triangulation step was critical to guarantee methodological soundness. For each candidate tactic, we conducted a targeted literature search using the tactic's conceptual keywords and related terminology. When prior work provided an analogous practice, pattern, or empirical observation, we cited it to support the tactic's plausibility and environmental relevance. When no supporting literature was found, the tactic was treated as a potentially novel contribution, provided that the code evidence and sustainability rationale were sufficiently strong. In cases where the literature search revealed contradictions or conceptual misalignment, the tactic was excluded. This systematic triangulation ensured that every tactic retained in the final catalog was grounded in (i) concrete code evidence, (ii) conceptual coherence, and (iii) alignment with current knowledge in sustainable software and ML engineering.} The process to reproduce our work is in the online appendix \cite{appendix}.



\subsection{Threats to validity}
\label{sec:threats}
Multiple factors may have influenced the study's conclusions and its generalizability. To ensure the robustness of our research design, we identified and addressed potential threats to validity during the study's design phase. In this section, we outline the threats that could have affected the study's results and describe the mitigation strategies incorporated before and during its execution.

\smallskip
\textbf{Construct Validity.} Construct validity assesses how well the studied variables and metrics represent the intended theoretical constructs~\cite{wohlin2012experimentation}. \revised{In our work, we relied on the catalog of green architectural tactics defined by Järvenpää et al.~\cite{10.1145/3639475.3640111}. While this catalog is the most comprehensive taxonomy available, some tactics, e.g., \textsl{`Clean and Standardize Data'} or \textsl{`Consider Use of Data Augmentation'}, also correspond to common ML best practices aimed at improving model performance. Their inclusion as green tactics follows the definitions provided in the catalog rather than an independent re-evaluation of their environmental impact. This overlap may blur the conceptual boundary between sustainability-oriented practices and general ML development heuristics. We acknowledge this limitation and see future work as an opportunity to refine and empirically reassess the conceptual scope of green architectural tactics, particularly regarding their distinction from standard ML practices.}

LLMs can produce errors if prompts do not accurately reflect task requirements, leading to incorrect outputs. To address this, the first two authors iteratively refined the prompt to align with the research questions and incorporate sustainable tactics \newchanges{using PRIMES framework \cite{de2024framework}}. Misinterpretation of tactic definitions also posed a risk, which we mitigated by using well-established definitions from the literature \cite{10.1145/3639475.3640111}. 
\newchanges{When new tactics emerged from Claude 3 Haiku, the first author reviewed them with the second author to ensure that only relevant tactics were considered. To further reduce the impact of hallucinations and improve the accuracy of tactics extraction, a manual validation phase was conducted. This process ensured that only well-founded and actionable tactics were included in the final analysis. Although this semi-automated process may limit the overall recall of tactics, we prioritized accuracy to ensure that the reported information was correct. We recognize that other green tactics may exist, and future work may explore broader methods of retrieval.}

More generally, the data analysis was conducted by the first two authors, who have substantial expertise in AI software engineering. The first author, a Ph.D. student in ML and SE4AI, has four years of experience in ML model development and three years in ML sustainability. The second author, an associate professor, has over 15 years of experience in ML and SE4AI.

\smallskip
\textbf{Internal Validity.} Internal validity assesses if the results are due to the studied variables rather than confounding factors~\cite{wohlin2012experimentation}. In our model evaluation, potential threats arose from the randomness of LLM outputs, leading to varying results even with the same prompt. Incorrectly formulated prompts could result in incomplete/inaccurate extraction of green tactics. In this respect, we refined the prompts iteratively, ran preliminary tests, and held meetings to improve accuracy.
As part of these preliminary tests, we evaluated the determinism of the LLMs by assessing consistency across runs and comparing the outputs with a manually constructed oracle. Based on these results, we selected the prompt that offered the best balance of determinism and informativeness. \revised{Although we fixed all sampling parameters to reduce non-determinism, slight variations may still occur due to model updates or infrastructure differences. Future work may investigate fully deterministic inference pipelines or calibration procedures to further limit this variability.} Despite this effort, we recognize that LLMs are inherently non-deterministic. Nevertheless, we accept this limitation in exchange for the models' strong pattern-recognition capabilities and effectiveness in tactical identification. 

\minor{An additional threat to internal validity concerns the choice to process source code at the file level without dividing it into chunks. Our pipeline assumes that individual Python files fall within the contextual window of the evaluated LLMs and does not record per-file token statistics. Although this design choice aims to preserve inter-file dependencies and avoid selective analysis, alternative strategies, such as chunking or hierarchical analysis, could expose different contextual information to the model and potentially yield different classification results. As a result, the detected presence of certain green architectural tactics may vary with the processing strategy employed.}

\revised{A further potential threat relates to the choice of relying on a single LLM, rather than an ensemble of LLMs, for the large-scale extraction of green architectural tactics. In this respect, the dependence on a single model may introduce model-specific biases. To mitigate this concern, we conducted preliminary experiments comparing multiple LLMs; Claude 3 Haiku consistently achieved the highest accuracy, provided richer explanatory outputs, and demonstrated the strongest capacity to uncover previously undocumented tactics. These characteristics were essential for addressing the study's research questions, making it the most reliable and cost-effective model for full-scale analysis. At the same time, adopting an ensemble of heterogeneous LLMs would have introduced substantial methodological and practical challenges. Ensembles require reconciling divergent reasoning styles, output formats, and conflict cases, which would reduce interpretability and complicate tracing predictions back to clear decision paths. This is particularly critical in our study, where explanation quality and tactic justification are core requirements. Moreover, executing an ensemble across hundreds of repositories would have significantly increased both monetary and computational cost, creating feasibility issues and ultimately contradicting the sustainability objectives underpinning our research, as ensemble-based analysis inherently consumes more energy. For these reasons, we opted for a single-model configuration.}

Finally, while Claude 3 Haiku could identify code related to tactics, it often omitted specific code snippets, only indicating the tactic's presence. To mitigate this source of inaccuracy, we tracked the repository and file names to verify the tactic’s existence and location.

\smallskip
\textbf{External Validity.} External validity assesses how well our findings can be generalized beyond the specific research context~\cite{wohlin2012experimentation}. A primary concern is the programming language used. \minor{Our study focuses on \textsc{Python} files, given its widespread use in ML and frequent appearance in existing datasets. While other languages are used in ML, we prioritized \textsc{Python} due to its global adoption by practitioners, and it is the dominant language for ML development and supported by most contemporary LLMs; however, green architectural tactics may manifest differently in systems implemented in other languages (e.g., C++, Java), potentially affecting both their detectability and prevalence. Similarly, runtime configurations, such as hardware characteristics, library versions, and frameworks, as well as implementation practices, can influence how sustainable practices are implemented in practice. Although these factors may limit generalization outside the studied context, the identified tactics and observed trends remain relevant as conceptual and architectural guidance for ML-based systems in Python.}

Another concern relates to dataset quality and generalizability. To mitigate this issue, we selected two well-established datasets, \textsc{NICHE} \cite{widyasari2023niche} and the dataset by Gonzalez \etal \cite{gonzalez2020state}, and applied preprocessing to remove outdated or irrelevant projects. Finally, we chose sustainable tactics based on Järvenpää et al. \cite{10.1145/3639475.3640111}, selecting clusters suitable for code-level analysis, a key requirement for our methodology. \revised{More specifically, the scope of our analysis is limited to green architectural tactics whose implementation leaves observable and unambiguous code-level footprints. Practices that require dataset inspection, deployment metadata, infrastructure-level configurations, or runtime information could not be analyzed by static code analysis and were therefore excluded. This boundary does not diminish the relevance of these tactics but constrains the generalizability of our findings to code-detectable sustainability practices.} Future research may explore the relevance of tactics excluded from this study. Additionally, we used default parameters for both Claude 3 Haiku and GPT-4o to reflect typical model performance, ensuring a standardized and replicable evaluation.

\smallskip
\textbf{Conclusion Validity.} Conclusion validity concerns the reliability of conclusions drawn from the data~\cite{wohlin2012experimentation}. In \textbf{RQ$_1$}, we evaluated classifier performance based on a sizable sample, but we cannot fully generalize these results to the entire ML project population. Similarly, the new tactics found in \textbf{RQ$_2$} may not apply universally across all ML projects. Another potential threat arises from using different LLMs for extracting sustainable tactics. We used GPT-4o \cite{gptkey} and Claude \cite{haikukey}, which differ in architecture and training data, and conducted a comparative evaluation, measuring their accuracy against a manually constructed oracle to select the most reliable model for accuracy and tactic discovery. Throughout our study, we followed the PRIMES framework \cite{de2024framework} to guide the research process and ensure methodological rigor. \revised{Finally, we note that although our study assesses the adoption and emergence of green architectural tactics, it does not quantify their actual energy or carbon impacts. Measuring the magnitude of these effects requires controlled experimentation and runtime profiling, which fall outside the scope of static code analysis. Future work will specifically focus on evaluating the energy and carbon savings associated with these tactics to complement the findings presented here.}

%% file: 5_Result.tex
\section{Analysis of the Results}
\label{sec:result}
In this section, we report quantitative and qualitative insights from the repository mining study to address the \textbf{RQ}s.

\subsection{\textbf{RQ\textsubscript{1}.} Adoption of Green Tactics in Practice}
\begin{figure}[h]
    \centering
    \includegraphics[width=1\linewidth]{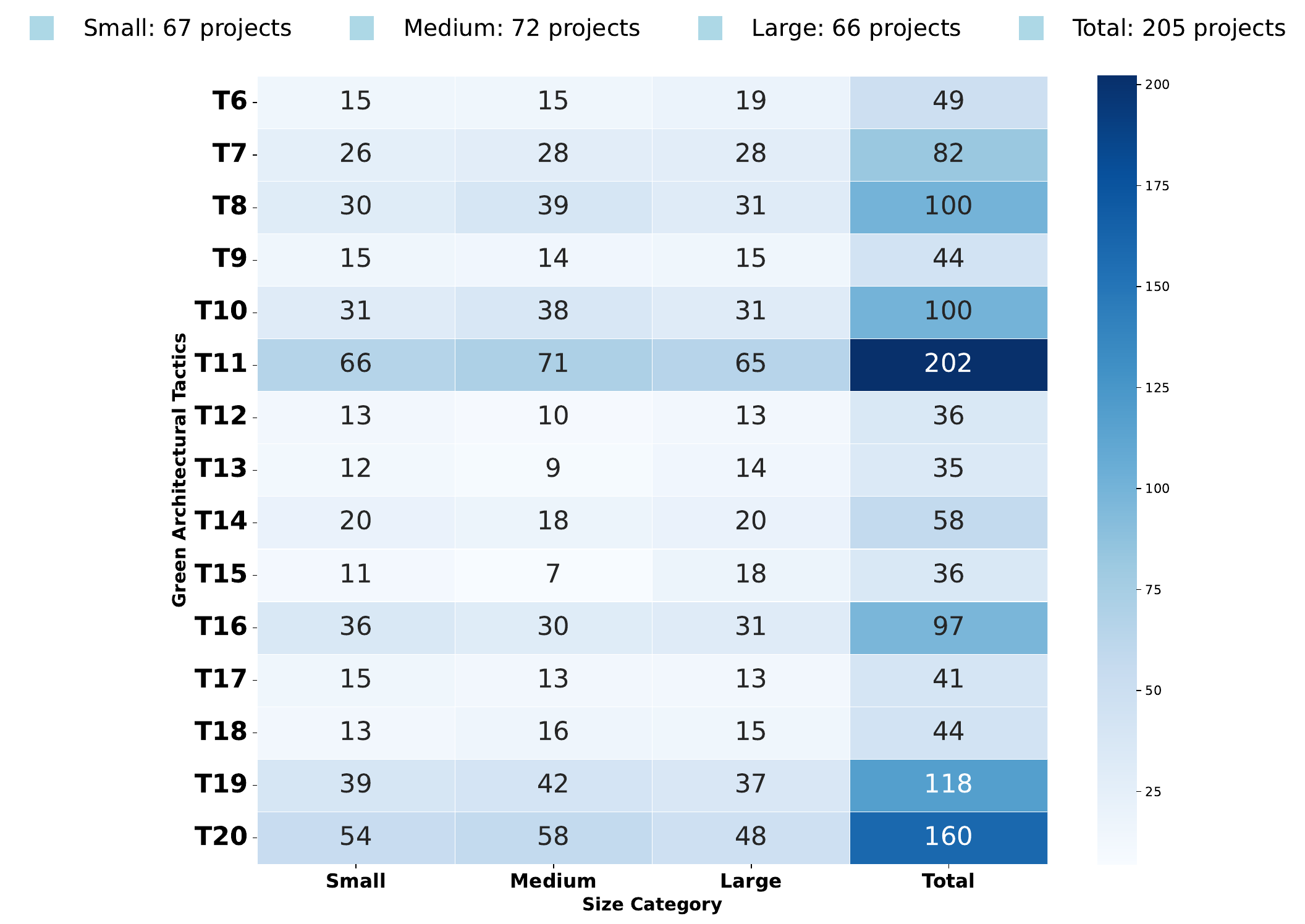}
    \caption{\revised{Frequency of Green Tactics in ML Projects.}}
    \label{fig:frequency}
\end{figure}

\revised{Figure \ref{fig:frequency} shows the frequency of adoption of green tactics in the considered 205 ML projects. It shows the number of projects that implemented each specific tactic, offering insights into the popularity of these strategies within real software projects across four size categories: \textsl{`Small'}, \textsl{`Medium'}, \textsl{`Large'}, and the full set of projects. Figure \ref{fig:frequency} highlights that adoption varies substantially across tactics but remains broadly comparable across size categories, with no single size group dominating the adoption of all tactics.} 

\revised{The heatmap shows that some tactics are consistently adopted across all size groups. Several green architectural tactics show consistently high adoption across projects of different sizes.For instance, \textsl{T11} (\textsl{`Use built-in library functions'}) is almost universally adopted, appearing in 66 small (98.5\%), 71 medium (98.6\%), and 65 large projects (97.0\%), for a total of 202 projects (98.5\%). A second group of tactics exhibits moderate adoption levels. Notably, \textsl{T8} (\textsl{`Decrease model complexity'}) and \textsl{T10} (\textsl{`Use dynamic parameter adaptation'}) display nearly identical adoption patterns across project sizes: each is implemented by 31 small (44.8\%), 39 medium (54.2\%), and 31 large projects (46.3\%), corresponding to 100 projects overall (48.8\%). Similarly, \textsl{T14} (\textsl{`Design for memory constraints'}) appears in 20 small (29.9\%), 18 medium (25.0\%), and 20 large projects (29.9\%), for a total of 58 projects (28.3\%).\textsl{T19} also shows relatively widespread adoption, being present in 39 small (58.2\%), 42 medium (58.3\%), and 37 large projects (55.2\%), amounting to 118 projects (57.6\%) overall. Conversely, several tactics exhibit lower adoption rates across all size categories. For example, \textsl{T12} is adopted by 10 small (14.9\%), 12 medium (16.7\%), and 13 large projects (19.4\%), totaling 35 projects (17.1\%).}

\revised{Similarly, \textsl{T13} appears in 14 small (20.9\%), 12 medium (16.7\%), and 10 large projects (14.9\%), for a total of 36 projects (17.6\%). \textsl{T9} shows comparably limited adoption, with 11 small (16.4\%), 19 medium (26.4\%), and 14 large projects (20.9\%), amounting to 44 projects (21.5\%). The remaining tactics show heterogeneous adoption levels. In particular, \textsl{T6} is present in 49 projects (23.9\%), \textsl{T7} in 82 projects (40.0\%), \textsl{T15} in 36 projects (17.6\%), \textsl{T16} in 97 projects (47.3\%), and \textsl{T20} in 160 projects (78.0\%), with varying distributions across project size categories.}

\revised{Our results show relatively comparable adoption rates across project size categories, we note a slight tendency for some tactics to appear less frequently in large projects. We do not interpret this as evidence that larger systems empirically integrate green tactics less often; rather, it may reflect general observations in software engineering literature that larger and more mature codebases typically involve greater architectural complexity, more dependencies, and higher coordination overhead~\cite{cataldo2006identification}. These factors have been reported to make the introduction of architectural changes more challenging in general. We therefore present this only as a potential interpretation informed by prior studies on software evolution, not as a conclusion derived from our data.}

\begin{table}[h]
\caption{\revised{P-values of logistic regression models assessing the relationship between project characteristics and tactic adoption.}}
\label{table:tactics_metrics}
\centering
\footnotesize
\begin{tabular}{|c|c|c|c|c|c|c|}
\rowcolor{black}
\textcolor{white}{\textbf{Tactic}} &
\textcolor{white}{\textbf{Commits}} &
\textcolor{white}{\textbf{Contributors}} &
\textcolor{white}{\textbf{Forks}} &
\textcolor{white}{\textbf{Age}} &
\textcolor{white}{\textbf{Size}} &
\textcolor{white}{\textbf{Stars}} \\
\hline
\rowcolor{gray10} T6  & 0.209 & 0.477 & 0.051 & \cellcolor[HTML]{C0BDDA} 0.007 & 0.114 & 0.058 \\
\hline
T7  & 0.794 & 0.713 & 0.322 & 0.466 & 0.228 & 0.322 \\
\hline
\rowcolor{gray10} T8  & 0.819 & 0.918 & 0.158 & 0.520 & 0.301 & 0.127 \\
\hline
T9  & 0.703 & 0.143 & 0.540 & 0.063 & 0.300 & 0.970 \\
\hline
\rowcolor{gray10} T10 & 0.854 & 0.266 & 0.400 & \cellcolor[HTML]{C0BDDA} 0.050 & 0.159 & 0.517 \\
\hline
T11 & 0.792 & 0.636 & 0.951 & 0.877 & 0.126 & 0.609 \\
\hline
\rowcolor{gray10} T12 & 0.351 & 0.507 & 0.114 & \cellcolor[HTML]{C0BDDA} 0.003 & 0.663 & 0.137 \\
\hline
T13 & 0.155 & 0.502 & 0.066 & \cellcolor[HTML]{C0BDDA} 0.006 & 0.962 & 0.094 \\
\hline
\rowcolor{gray10} T14 & 0.164 & 0.111 & \cellcolor[HTML]{C0BDDA} 0.046 & \cellcolor[HTML]{C0BDDA} 0.004 & 0.161 & 0.072 \\
\hline
T15 & 0.450 & 0.712 & 0.172 & \cellcolor[HTML]{C0BDDA} 0.008 & 0.566 & 0.183 \\
\hline
\rowcolor{gray10} T16 & 0.412 & 0.628 & 0.066 & 0.541 & 0.676 & \cellcolor[HTML]{C0BDDA} 0.047 \\
\hline
T17 & 0.549 & 0.409 & 0.214 & \cellcolor[HTML]{C0BDDA} 0.034 & 0.717 & 0.253 \\
\hline
\rowcolor{gray10} T18 & 0.771 & 0.420 & 0.289 & \cellcolor[HTML]{C0BDDA} 0.044 & 0.454 & 0.350 \\
\hline
T19 & 0.787 & 0.200 & 0.093 & 0.796 & 0.616 & \cellcolor[HTML]{C0BDDA} 0.029 \\
\hline
\rowcolor{gray10} T20 & 0.781 & \cellcolor[HTML]{C0BDDA} 0.030 & 0.879 & \cellcolor[HTML]{C0BDDA} 0.073 & 0.693 & 0.518 \\
\hline
\end{tabular}
\end{table}
\revised{Table \ref{table:tactics_metrics} reports the p-values obtained from logistic regression models evaluating whether project characteristics are statistically associated with the adoption of each tactic. The table shows which predictors reach conventional significance thresholds for each tactic. Across tactics, Project Age shows statistically significant p-values (p $<$ 0.05) for several tactics, including T6, T12, T13, T14, T15, T17, and T18. Other predictors show significance in specific cases: for example, Forks is significant for T14, Stars for T16 and T19, Contributors is significant for T20, and Size generally does not reach significance for any tactic. Most predictors present p-values above 0.05 for most tactics, indicating no statistically detectable relationship under the fitted logistic models.}

\rqanswer{1}{Green tactics proposed in the literature are widely adopted in ML projects. However, some tactics are less commonly adopted, highlighting the need to raise awareness and provide better tool support to facilitate their broader integration.}

\begin{table*}[h]
    \caption{New Green Tactics for ML-Enabled Systems. (*) Indicate energy efficiency is a secondary quality assurance (QA)}
    \label{table:list_new_tactics}
    \centering  
    \footnotesize
    \rowcolors{1}{gray!15}{white}
    \begin{tabular}{|p{65pt}|p{80pt}|p{65pt}|c|c|}
    \hline
    \rowcolor{black}
    \textcolor{white}{\textbf{Cluster}} & \textcolor{white}{\textbf{Tactic}} & \textcolor{white}{\textbf{Target QA}} & \textcolor{white}{\textbf{\#Projects}} & \textcolor{white}{\textbf{Source}} \\
    \hline
    Data-Centric & T31: Clean and Standardize Data & Performance* &  \revised{9	(4.39\%)} & \cite{zha2023data,de2025sustainability} \\
    \cline{2-5}
    & T32: Consider Use Data Augmentation & Accuracy* &  \revised{24	(11.71\%)} & \cite{salehi2023data,shorten2019survey,de2025sustainability} \\
    \hline
    Model Optimization & T33: Ensure Gradient Stability & Performance* &  \revised{5 (2.44\%)} & \cite{zhang2019gradient} \\
    \cline{2-5}
    & T34: Use Adaptive Learning Rate Scheduling & Energy efficiency & \revised{12 (5.85\%)} & \cite{smith2017cyclical,wang2023empirical}\\
    \hline
    Model Training & T35: Employ Parallel Training & Performance* & \revised{28	(13.66\%)} & \cite{dean2012large,huang2019gpipe}\\
    \cline{2-5}
    & T36: Use Early Stopping & Energy efficiency & \revised{88	(42.93\%)} & \cite{reguero2024energy,li2020gradient} \\
    \hline
    Deployment & 
    T37: Use Deployment-Specific Model Export Strategies & Energy efficiency &  \revised{6	(2.93\%)} & \cite{hampau2022empirical}\\
    \hline
    Management & T38: Consider Model Reproducibility & Recoverability* & \revised{7	(3.41\%)} & \cite{pan2022decomposing,qi2023reusing} \\
    \cline{2-5}
    & T39: Dynamic Resource Management & Resource utilization* & \revised{17 (8.29\%)} & \cite{chiang2023dynamic}\\
    \hline
    \end{tabular}
\end{table*}

\subsection{\textbf{RQ\textsubscript{2}.} Elicitation of Additional Green Tactics}
Table \ref{table:list_new_tactics} catalogs the set of new tactics identified in \textbf{RQ$_2$}, along with their frequency in the ML projects considered in the study. As shown, the catalog contains nine green architectural tactics for ML-enabled systems, outlined below. 
\revised{Each tactic has been elicited through a semi-automated analysis supported by an LLM (Claude 3 Haiku), followed by a manual validation and refinement process aimed at ensuring precision and relevance (see Section \ref{sec:extraction} for details on the validation and triangulation procedure).}
In this context, a tactic is included only if it represents a high-level design decision that contributes to energy or resource efficiency and exhibits architectural scope, consistent with the definition adopted in prior work~\cite{10.1145/3639475.3640111}. To enhance clarity and facilitate traceability, each tactic is described following a common structured template composed of three elements:
\begin{itemize}
    \item \newchanges{\textsl{\underline{Definition}}: which explains the core idea and scope of the tactic;}
    \item \newchanges{\textsl{\underline{Rationale and Sustainability Objectives}}: which outlines the expected sustainability benefits of the tactic, supported by empirical evidence or relevant literature;}
    \item \newchanges{\textsl{\underline{Code-example}}: which provides an implementation example illustrating the tactics adopted in real-world ML projects.}
\end{itemize}

\subsubsection{\faLeaf\ T31: Clean and Standardize Data}
\newchanges{\textsl{Definition:} this tactic involves transforming raw, unprocessed data into a clean, consistent, and structured format. It includes steps such as noise removal, error correction, handling missing values, resolving inconsistencies, and enforcing standard formats and conventions \cite{zha2023data,de2025sustainability}.}\\
\newchanges{\textsl{Rationale and Sustainability Objectives:}
as highlighted by Zha et al.~\cite{zha2023data}, clean and standardized data contribute to more efficient training by reducing the number of iterations required for convergence and minimizing the need for extensive hyperparameter tuning. These improvements can lead to lower computational demand and energy consumption, thus supporting sustainability objectives. However, if not carefully applied, it may negatively affect fairness by removing rare or minority instances, which are important for representative learning.}\\
\textsl{Code-example:} example of function to clean and standardize text by handling encoding issues \cite{t31}, removing line breaks, and unescaping HTML entities:
\begin{lstlisting}
def transform_clean_text(raw): 
    decoded = raw.encode("ISO-8859-1").decode("utf-8") 
    html_unescaped = Transformer.__html_parser.unescape(decoded)
    html_unescaped = re.sub(r"\r \n", " ",html_unescaped) 
    return html_unescaped 
\end{lstlisting}

\subsubsection{\faLeaf\ T32: Consider Use Data Augmentation} 

\newchanges{\textsl{Definition:} This tactic consists of synthetically expanding the training dataset by generating new examples from existing data. Common methods include transformations such as rotation, flipping, noise injection, or more advanced generative models (e.g., GANs or VAEs) \cite{salehi2023data,shorten2019survey,de2025sustainability}.}\\
\newchanges{\textsl{Rationale and Sustainability Objectives:} data augmentation improves model generalization by increasing data diversity without requiring new data collection or additional model complexity. According to Salehi and Schmeink~\cite{salehi2023data}, it enhances data efficiency and reduces energy consumption by lowering the number of training epochs needed to reach acceptable performance. This results in reduced compute time and fewer carbon emissions, making it an effective green strategy in data-constrained scenarios. However, it may introduce artifacts or noisy instances that negatively affect explainability or lead to unpredictable model behaviors if not well controlled. Careful calibration of augmentation techniques is therefore required.}\\
\textsl{Code-example:} a basic augmentation \cite{t32} for audio data by randomly adjusting gain or swapping audio channels:
\begin{lstlisting}
def _augment_gain(audio, low=0.25, high=1.25):
    gain = low + torch.rand(1) * (high - low) 
    return audio * gain  
def _augment_channelswap(audio):
    if audio.shape[0] == 2 and torch.FloatTensor(1).uniform_() < 0.5: 
        return torch.flip(audio, [0]) 
    return audio
\end{lstlisting}

\subsubsection{\faLeaf\ T33: Ensure Gradient Stability} 
\newchanges{\textsl{Definition:} this tactic involves applying techniques such as gradient clipping or normalization to maintain stable gradient values during model training. It aims to prevent vanishing or exploding gradients, which are particularly problematic in deep or recurrent neural networks.}\\
\newchanges{\textsl{Rationale and Sustainability Objectives:} unstable gradients can significantly slow convergence or cause divergence in deep learning models. Zhang et al.~\cite{zhang2019gradient} provide theoretical and empirical evidence that gradient clipping improves training stability and allows faster convergence by enabling the optimizer to traverse non-smooth regions more effectively. As faster convergence reduces the number of required training epochs, this tactic contributes to lowering energy consumption and computational cost. However, excessive clipping may hinder learning capacity or affect convergence to optimal minima, potentially impacting final model performance or explainability. Careful tuning is required to balance efficiency with learning quality.}\\
\textsl{Code-example:} gradient clipping \cite{t36}, which helps to stabilize the training process and prevent numerical issues:
\begin{lstlisting}
trainer = pl.Trainer(..., gradient_clip_val = conf["training"]["gradient_clipping"])
\end{lstlisting}

\subsubsection{\faLeaf\ T34: Use Adaptive Learning Rate Scheduling} 

\newchanges{\textsl{Definition:} this tactic involves dynamically adjusting the learning rate during training, often in response to validation performance or training epochs. Common strategies include decay schedules or cyclical learning rate (CLR) policies, where the learning rate oscillates between bounds to improve convergence and stability \cite{smith2017cyclical,wang2023empirical}.}\\
\newchanges{\textsl{Rationale and Sustainability Objectives:} adaptive learning rate schedules accelerate convergence and improve model quality without extensive manual tuning. Smith~\cite{smith2017cyclical} showed that cyclical schedules can achieve the same accuracy as fixed schedules in fewer iterations, reducing training cost. Wang et al.~\cite{wang2023empirical} found that CLR enables smaller batch sizes and faster convergence in neural machine translation, making it a practical choice in resource-constrained settings. These benefits contribute to reducing energy consumption and computational waste. However, poorly configured schedules might lead to instability or suboptimal convergence, affecting reproducibility and accuracy. Careful boundary selection is required.}\\
\textsl{Code-example:} example of learning rate scheduling configuration for training \cite{t37}:
\begin{lstlisting}
lr_config = dict(policy = 'step', warmup='linear', warmup_iters = 500, warmup_ratio = 1.0 / 3, ...)
\end{lstlisting}

\subsubsection{\faLeaf\ T35: Employ Parallel Training} 
\newchanges{\textsl{Definition:} this tactic distributes the training of a machine learning model across multiple processing units (e.g., GPUs, TPUs, CPUs, or nodes in a cluster) to exploit hardware parallelism. Strategies include data parallelism, model parallelism, and pipeline parallelism \cite{dean2012large,huang2019gpipe}.}\\
\newchanges{\textsl{Rationale and Sustainability Objectives:} parallel training accelerates convergence by processing larger datasets or deeper models faster. Dean et al.\cite{dean2012large} showed that their DistBelief framework achieved up to 12$\times$ speedup using 81 machines, drastically reducing training time and enabling the use of larger models within a fixed budget. Huang et al.\cite{huang2019gpipe} introduced GPipe, demonstrating almost linear throughput scaling via pipeline parallelism, allowing models up to 83.9B parameters to be trained efficiently. Faster convergence and higher hardware utilization reduce wall-clock training time and energy consumption. However, it may increase implementation complexity and resource cost, and introduce reproducibility challenges due to hardware-specific behavior and synchronization differences. Robust orchestration is required to preserve training correctness.}\\
\textsl{Code-example:} example of distributed training \cite{t38}:
\begin{lstlisting}
from tensorflow.python.distribute.cross_device_ops import AllReduceCrossDeviceOps

cross_tower_ops = AllReduceCrossDeviceOps('nccl', num_packs=n_gpus)
strategy = tf.distribute.MirroredStrategy(
    devices=[f'/gpu:{i}' for i in range(n_gpus)],
    cross_device_ops=cross_tower_ops)
\end{lstlisting}

\subsubsection{\faLeaf\ T36: Use Early Stopping} 
\newchanges{\textsl{Definition:} this tactic involves monitoring the model's performance on a validation set and halting training once the performance ceases to improve. The goal is to prevent overfitting and ensure the model is trained only for the necessary number of epochs.}\\
\revised{\textsl{Rationale and Sustainability Objectives:} early stopping reduces unnecessary training by avoiding overfitting phases, directly lowering computational cost and energy use. During neural network training, it helps prevent the memorization of this label noise, which can cause overfitting and result in poor generalization performance. Additionally, early stopping reduces both training duration and computational load, leading to significant energy savings during the training of neural network. This tactic minimizes redundant computation and offers one of the clearest energy-performance trade-offs in green AI \cite{li2020gradient,reguero2024energy}. However, it may increase implementation complexity and resource cost, and introduce reproducibility challenges due to hardware-specific behavior and synchronization differences. Robust orchestration is required to preserve training correctness.}\\
\textsl{Code-example:} example of early stopping \cite{t39}:
\begin{lstlisting}
from pytorch_lightning.callbacks import EarlyStopping  
early_stopping = EarlyStopping(monitor=monitor, mode=direction, min_delta=0.0, patience=100, strict=True, verbose=False, check_finite=True)
\end{lstlisting}

\subsubsection{\faLeaf\ T37: Use Deployment-Specific Model Export Strategies} 
\newchanges{\textsl{Definition:} this tactic involves exporting trained models into optimized formats (e.g., ONNX, WebAssembly) that are tailored to the characteristics of the deployment environment, such as edge devices or embedded systems with constrained memory or processing power. The export format is chosen to minimize runtime overhead while maximizing compatibility and performance.}\\
\newchanges{\textsl{Rationale and Sustainability Objectives:} deployment-specific export strategies optimize the inference phase by reducing overhead and improving runtime efficiency. Hampau et al.~\cite{hampau2022empirical} conducted an empirical comparison of ONNX, Docker, and WebAssembly (WASM) deployments across multiple edge devices. Their results show that ONNX runtime offers the best trade-off between energy consumption and execution time, while WASM achieves the lowest memory footprint. These findings confirm that selecting the appropriate export format based on target hardware can significantly reduce energy usage and resource consumption in real-world deployments. However, the conversion process may introduce compatibility issues or require additional validation steps to ensure correctness, potentially affecting reliability or reproducibility if not well-managed.}\\
\textsl{Code-example:} converting a model in ONNX \cite{t40}:
\begin{lstlisting}
torch_onnx_export(decoder, args = tuple(input_list), export_params = True, input_names = input_names, output_names = output_names, dynamic_axes = dynamic_axes, opset_version = 12, do_constant_folding = True, verbose = verbose, ...)
\end{lstlisting}

\subsubsection{\faLeaf\ T38: Consider Model Reproducibility} 
\newchanges{\textsl{Definition:} this tactic focuses on ensuring that ML systems can be reproduced under consistent conditions by explicitly controlling randomness, logging configurations, and structuring code for repeatability. Common practices include fixing random seeds, recording hyperparameters, tracking datasets and code versions, and modularizing model components \cite{pan2022decomposing,qi2023reusing}.}\\
\newchanges{\textsl{Rationale and Sustainability Objectives:} improving reproducibility reduces the need for redundant experiments and trial-and-error iterations, thus saving time and computational resources. Qi et al.\cite{qi2023reusing} show that re-engineering models into modular, reusable components can reduce inference time by 42.41\% and weight usage by 89.89\%, decreasing overhead in repeated experiments. Similarly, Pan and Rajan\cite{pan2022decomposing} report that decomposing CNNs into reusable modules lowers CO$_2$ emissions by up to 37$\times$ compared to training from scratch.}\\
\textsl{Code-example:} example for track training progress \cite{t41}:
\begin{lstlisting}
wandb = dict(init_kwargs = dict(project = "caltech", name = "convmlp_2x16", entity = "mlp", config = dict(work_dirs = "${work_dir}", total_step = "${runner.max_epochs}")), interval = 50) 
\end{lstlisting}

\subsubsection{\faLeaf\ T39: Dynamic Resource Management} 
\newchanges{\textsl{Definition:} this tactic refers to the dynamic allocation and adjustment of computational resources (e.g., CPU, GPU, memory) during model training and inference, based on real-time workload demands. It includes techniques such as auto-scaling, job preemption, elastic allocation, and topology-aware scheduling to optimize resource use throughout the ML pipeline \cite{chiang2023dynamic}.}\\
\newchanges{\textsl{Rationale and Sustainability Objectives:} Chiang et al.~\cite{chiang2023dynamic} demonstrate that dynamic resource management in ML pipelines significantly improves efficiency: their system, DynamoML, achieved a 60\% increase in system throughput and a 70\% reduction in training time compared to standard Kubernetes. These improvements reduce energy consumption by avoiding idle resource use and minimizing job execution time. Techniques like elastic GPU sharing and workload-aware scheduling contribute directly to energy savings and more sustainable ML operations. However, it may introduce complexity in orchestration, impact reproducibility due to resource variability, and require additional infrastructure (e.g., schedulers, controllers) to ensure consistent behavior across dynamic conditions.}\\
\textsl{Code-example:} a shared memory management or efficient data management between processes or threads \cite{t42}:
\begin{lstlisting}
sa_array = sa.create(name, data.shape, data.dtype) np.copyto(sa_array, data)
\end{lstlisting}

\rqanswer{2}{We extracted nine new green tactics adopted by developers. Among them, \textsl{`Use Early Stopping'} and \textsl{`Employ Parallel Training'} are the most adopted.} 

%% file: 6_Discussion.tex
\section{Discussion and Implications}
\label{sec:discussion}
\begin{figure*}[ht]
    \centering
    \includegraphics[width=1\linewidth]{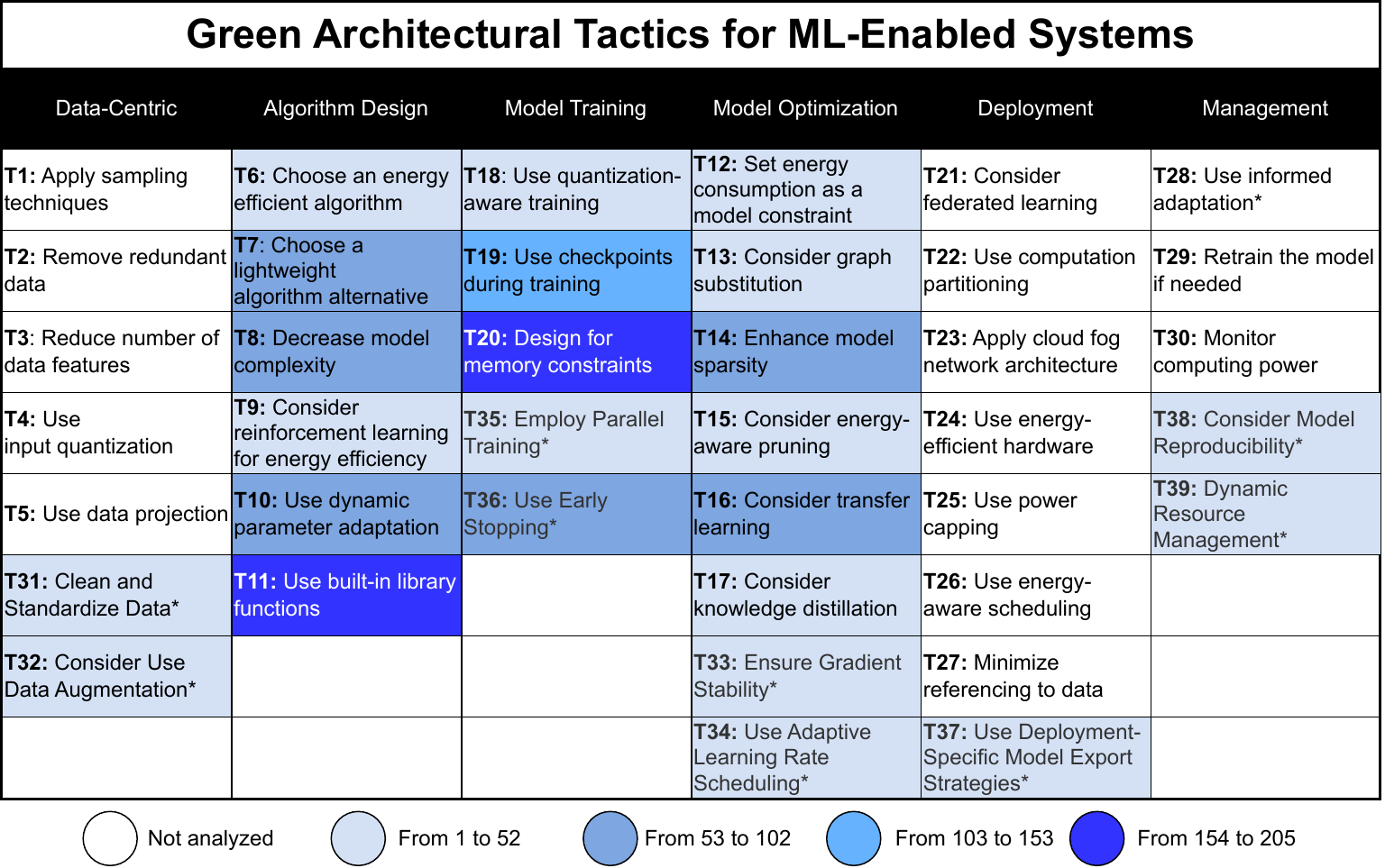}
    
    \caption{\revised{Catalog of 39 Green Architectural Tactics for ML-Enabled Systems, adopted from \cite{10.1145/3639475.3640111} and enhanced with use frequency of all tactics and the elicitation of nine new tactics. The symbol \textsl{`*'} indicates a new tactic found by our LLM-based mechanism. Colors indicate frequency using quartiles on 205 ML projects.}}
    \label{fig:new_catalog}
\end{figure*}

Our study highlights several implications for the SE community and for society by minimizing the energy footprint.

\smallskip
\textbf{On the Adoption of Green Tactics.} When comparing the adoption of green tactics identified in previous studies, particularly the catalog of 30 tactics by Järvenpää et al. \cite{10.1145/3639475.3640111}, we analyzed \revised{205 open-source ML projects} on GitHub using LLMs. Our analysis confirms that practitioners adopt many of these tactics. Specifically, tactics like \textsl{`Use of built-in library functions'} and \textsl{`Design for Memory Constraints'} have high adoption rates, suggesting that these tactics are well-supported and effectively integrated into current development processes. Conversely, other tactics such as \textsl{`Set Energy Consumption as a Model Constraint'} and \textsl{`Consider Graph Substitution'} exhibit lower adoption rates in our study. 
Although our analysis is based on open-source projects, the results have broader implications for different ML contexts. 
\minor{The quantitative analysis shows that some green architectural tactics are adopted consistently across projects, whereas others occur only sporadically, pointing to uneven levels of awareness or maturity in sustainable design practices. Notably, several of the tactics surfaced by the LLMs align with areas exhibiting low observed adoption, suggesting that these practices are not yet  documented or commonly considered during development. From a professional perspective, this report helps contextualize the findings, where frequently adopted tactics can be interpreted as established practices, while low-frequency and newly identified tactics represent valid but under-explored design options that professionals can consider when seeking to improve the sustainability of ML-based systems.}

Many tactics are applicable to proprietary and enterprise-level systems, but some, such as energy-aware optimizations, may require adaptations for specific domains such as resource-constrained environments. Green tactics also hold potential for improving accessibility and affordability by reducing hardware and energy costs, but their adoption must be balanced against potential trade-offs in performance and usability, particularly in critical applications such as healthcare and autonomous systems. Adoption barriers, including organizational resistance, lack of awareness, and perceived costs, are particularly evident in larger projects where coordination and competing priorities complicate sustainability efforts. Addressing these barriers will require better education, improved tooling, and incentives for adopting green tactics. Engaging directly with developers through interviews or analyzing project repositories could uncover additional insights into these challenges and inform strategies for overcoming them.

\impl{1}{Our results show that the adoption of green architectural tactics varies across different projects. This suggests a need for targeted strategies and tools tailored to address specific barriers. Future research should aim to explore the socio-technical reasons behind current adoption limitations.}

\medskip
\textbf{On the Elicitation of Additional Green Tactics.} When comparing the newly identified green tactics with those cataloged in existing literature, we find that developers are employing strategies not yet formally cataloged. This discovery has significant implications for both researchers and practitioners. The addition of these tactics expands the existing body of knowledge on green tactics \cite{10.1145/3639475.3640111,verdecchia2023systematic}, increasing the total to 39 tactics. The new catalog is shown in Figure \ref{fig:new_catalog}. The newly identified green tactics are marked with a \textsl{`*'} symbol. Furthermore, the figure illustrates the extent to which all the tactics analyzed in this study, both those previously elicited and the new ones, are employed within the systems considered. This updated catalog represents the main contribution of the work, (1) expanding the body of knowledge on green tactics by identifying previously unrecognized strategies and (2) providing insights into their practical adoption and prevalence within the systems analyzed. 

Notably, some of these new tactics, such as \textsl{`Use Early Stopping'}, exhibit high adoption rates (found in \revised{88} projects), suggesting their practical relevance and effectiveness. \revised{Indeed, early stopping (\textit{T36}) is beneficial when working with datasets that have lower-quality annotations, including mistakes or label noise. During neural network training, it helps prevent the memorization of this label noise, which can cause overfitting and result in poor generalization performance \cite{bai2021understanding}. Additionally, early stopping reduces both training duration and computational load, leading to significant energy savings during the training of neural networks \cite{reguero2024energy}. This sustainable tactic (\textit{T36}) can be implemented to lower energy consumption by either considering the entire neural network or focusing progressively on specific layers. Research shows that applying early stopping at later layers often has a greater impact \cite{bai2021understanding}.} This opens new avenues for further research into the impact and efficiency of these tactics. Empirical studies could focus on quantifying the environmental benefits and energy savings of these tactics in different ML systems. There is also a need to explore how these tactics can be adapted to proprietary or large-scale industrial systems, which may face unique challenges. Examining the interaction between green tactics and other software QA, such as scalability, reliability, and maintainability, could provide insights into their broader applicability and impact in ML-enabled systems. While our study emphasizes energy efficiency, other QAs also affect accuracy. Typically, achieving higher accuracy is viewed as a trade-off with energy consumption, where more training leads to increased energy use. A significant drop in accuracy to meet energy goals can make ML models unreliable for practical applications, underscoring the need to prioritize accuracy. However, recent research suggests that it is possible to reduce energy consumption without compromising accuracy \cite{10.1145/3639475.3640111,verdecchia2023systematic}. By understanding how different QAs interact, we can design systems that effectively balance energy efficiency, accuracy, and other requirements. By providing practical code examples, we aim to facilitate the integration of these tactics into more projects, potentially fostering wider adoption of green architectural tactics. This effort helps developers create more efficient systems and motivates the software engineering community to further embrace and enhance the use of emerging techniques. \revised{Finally, future research could focus on tactics not considered in our study and extend the analysis beyond Python-based projects to identify their adoption in real-world systems implemented in other programming languages, such as Java, C++, or Scala, thereby improving the generalizability of the findings.} 

\impl{2}{The discovery of nine new green architectural tactics highlights ongoing opportunities in sustainable software engineering. These findings encourage further exploration of their energy savings and integration into software quality frameworks, paving the way for studies and practical guidelines for developers. Importantly, offering practical code snippets for these tactics significantly lowers the barriers to adoption, making it easier for developers to integrate them into real projects. By demonstrating how to implement these tactics, we aim to enhance their accessibility and encourage wider adoption, ultimately contributing to the development of more sustainable ML-enabled systems.}

\smallskip
\textbf{Broader Impact.} The results of our work have a broader impact, aligning closely with the principles of Green AI \cite{schwartz2020green} and SDG Goals\footnote{\url{https://www.globalgoals.org/goals/}.}. By emphasizing strategies to reduce resource consumption and CO\textsubscript{2} emissions, green architectural tactics contribute significantly to addressing the environmental impact of AI technologies while promoting efficient use of resources. These strategies are particularly relevant in sectors that prioritize energy efficiency and environmental responsibility. Balancing energy efficiency with performance and usability remains a major challenge. Sectors such as healthcare and autonomous systems often demand uncompromised performance and reliability, creating potential trade-offs in implementing green tactics. Furthermore, while green architectural tactics promise to reduce hardware costs and resource requirements, achieving this balance requires careful consideration to ensure accessibility and affordability. These societal benefits demonstrate the importance of adopting sustainable tactics in ML development and underscore the need to explore these trade-offs further to guide informed decision-making.

To fully realize these benefits, collaboration between researchers and practitioners is essential. Researchers can focus on developing robust benchmarks and frameworks to evaluate the effectiveness of green tactics, while practitioners can integrate these strategies into different application areas. Working together, these stakeholders can ensure that green tactics mitigate environmental impacts and drive broader social progress, promoting a more sustainable and responsible future for AI technologies.

\impl{3}{Our study highlights the need for interdisciplinary collaboration among researchers and practitioners to adopt the benefits of green tactics. This requires initiatives that combine technological advances with regulatory support, ensuring that green tactics lead to both environmental sustainability and societal benefits.}

%% file: 7_Conclusion.tex
\section{Conclusion and Future Work}
\label{sec:conclusion}
In this paper, we explore the adoption of green tactics in ML-enabled systems through a study of \revised{205} open-source ML projects. We assess their practical use and identify undocumented practices that may enhance sustainability. We also develop a novel LLM-based mechanism for identifying green architectural tactics in software repositories. Our work provides the following contributions:
\begin{enumerate}
    \item The introduction of a novel LLM tool to identify tactics;

    \smallskip
    \item An empirical analysis on the adoption of existing green tactics in practice;

    \smallskip
    \item An empirical analysis aimed at discovering new tactics to expand existing catalogs;

    \smallskip
    \item A public replication package with study materials \cite{appendix}.
\end{enumerate}
\revised{Future work will explore the impact of green architectural tactics on sustainability through longitudinal studies. It will also examine socio-technical factors, such as knowledge gaps and implementation challenges, while promoting lesser used tactics through improved documentation, tools, and case studies. Moreover, since this study focused on Python based ML projects due to the language’s prevalence and compatibility with major ML frameworks, future work will extend the analysis to other languages (e.g., Java, C++, Scala) to enhance the generalizability of the findings and capture sustainability practices across diverse programming environments. Finally, we will integrate the new tactics and their code example into the Archive of Awesome and Dark Tactics \footnote{\url{https://s2group.cs.vu.nl/AwesomeAndDarkTactics/catalog}}. These steps will help integrate green tactics more effectively into ML development, contributing to the creation of more sustainable systems.}
